\documentclass[ijds,nonblindrev]{informs-ijds}

\OneAndAHalfSpacedXI

\usepackage{bm}
\usepackage{diagbox}
\usepackage{adjustbox}
\usepackage{makecell}
\usepackage{subcaption}
\usepackage{textcomp}
\usepackage[ruled,vlined]{algorithm2e}

\usepackage{natbib}
 \bibpunct[, ]{(}{)}{,}{a}{}{,}%
 \def\bibfont{\small}%
\usepackage[pdftex,colorlinks,
            urlcolor = blue,
            linkcolor=blue,
            anchorcolor=blue,
            citecolor=blue
            ]{hyperref}

\TheoremsNumberedThrough     
\ECRepeatTheorems

\EquationsNumberedThrough    

\MANUSCRIPTNO{IJDS-****-****.**}

\begin{document}


\RUNAUTHOR{Wei et al.} 

\RUNTITLE{Physics-informed statistical modeling for wildfire aerosols propagation}

\TITLE{\large{Physics-Informed Statistical Modeling for Wildfire Aerosols Process Using Multi-Source Geostationary Satellite Remote-Sensing Data Streams}}

\ARTICLEAUTHORS{
    \AUTHOR{Guanzhou Wei}
    \AFF{Department of Industrial Engineering, University of Arkansas, Fayetteville, \EMAIL{gwei@uark.edu}} 
    \AUTHOR{Venkat Krishnan}
    \AFF{PA Consulting \EMAIL{Venkat.Krishnan@paconsulting.com}}
    \AUTHOR{Yu Xie}
    \AFF{National Renewable Energy Laboratory, \EMAIL{Yu.Xie@nrel.gov}}
    \AUTHOR{Manajit Sengupta}
    \AFF{National Renewable Energy Laboratory \EMAIL{Manajit.Sengupta@nrel.gov}}
    \AUTHOR{Yingchen Zhang}
    \AFF{Utilidata \EMAIL{}}
    \AUTHOR{Haitao Liao}
    \AFF{Department of Industrial Engineering, University of Arkansas, Fayetteville, \EMAIL{liao@uark.edu}}
    \AUTHOR{Xiao Liu\thanks{Corresponding author}}
    \AFF{Department of Industrial Engineering, University of Arkansas, Fayetteville, \EMAIL{XL027@uark.edu}}
}

\ABSTRACT{
Increasingly frequent wildfires significantly affect solar energy production as the atmospheric aerosols generated by wildfires diminish the incoming solar radiation to the earth. Atmospheric aerosols are measured by Aerosol Optical Depth (AOD), and AOD data streams can be retrieved and monitored by geostationary satellites. However, multi-source remote-sensing data streams often present heterogeneous characteristics, including different data missing rates, measurement errors, systematic biases, and so on. To accurately estimate and predict the underlying AOD propagation process, there exist practical needs and theoretical interests to propose a physics-informed statistical approach for modeling wildfire AOD propagation by simultaneously utilizing, or fusing, multi-source heterogeneous satellite remote-sensing data streams. Leveraging a spectral approach, the proposed approach integrates multi-source satellite data streams with a fundamental advection-diffusion equation that governs the AOD propagation process. A bias correction process is included in the statistical model to account for the bias of the physics model and the truncation error of the Fourier series. The proposed approach is applied to California wildfires AOD data streams obtained from the National Oceanic and Atmospheric Administration. Comprehensive numerical examples are provided to demonstrate the predictive capabilities and model interpretability of the proposed approach. Computer code has been made available on GitHub. 
}



\maketitle

%


\section{Introduction}
Statistical modeling of spatio-temporal data arises from a spectrum of scientific and engineering applications, including environmental and natural processes \citep{Stroud2010, Guinness, Liu2016, Wikle2019, Ezzat}, quality and reliability engineering \citep{Liu2018a, Yan, Fang}, medical informatics \citep{Yao, Yang}, and so on. In this paper, we propose a physics-informed statistical  approach for modeling heterogeneous multi-source remote-sensing data streams from wildfire smoke propagation processes.

\subsection{Motivating Application}
\begin{figure}[h!]
  \centering
	\includegraphics[width=0.8\textwidth]{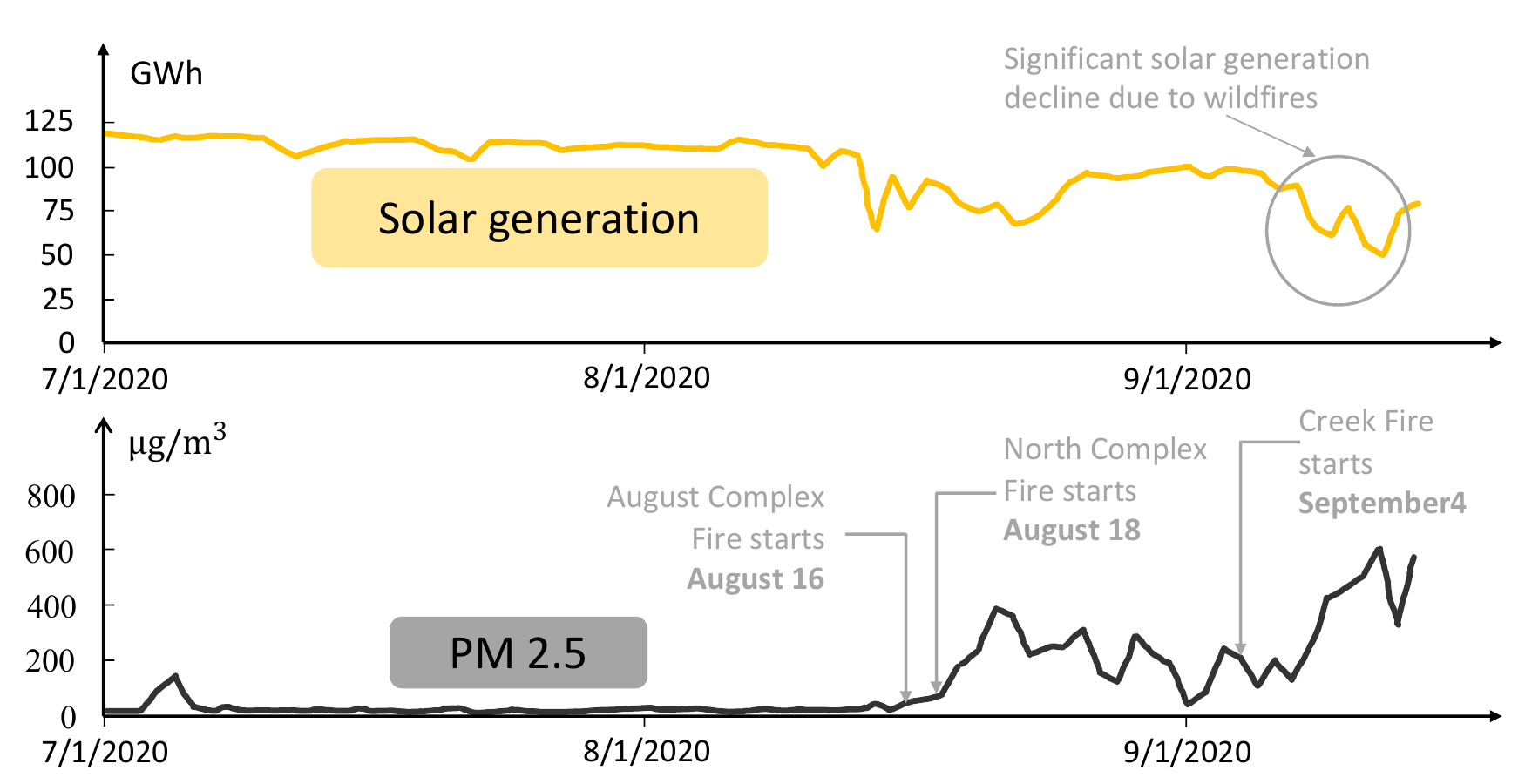}
	\caption{Daily California Independent System Operator solar generation (above) and peak air particulate matter level (bottom) \citep{York}}\label{fig:motivation}
  \end{figure}

The increasingly severe wildfires have significantly affected solar energy production in the U.S. over recent years. Although the installed solar generating capacity in California (CA) was increased by 5.3\% from September 2019 to June 2020, solar power generation in the first two weeks of September 2020 was 13.4\% lower than at the same time a year ago. Atmospheric aerosols generated by wildfires blocked the incoming solar radiation and were deposited on solar panel surface, reducing the received solar energy from photovoltaics.
In atmospheric science, aerosols are solid or semi-solid particles, such as PM$_{2.5}$ in the atmosphere, and produced by environmental disasters including volcanic eruptions, biomass burning, wildfires, etc. As shown in Figure \ref{fig:motivation}, the average daily solar generation in the CA Independent System Operator (which covers 90\% of utility-scale solar capacity in CA) declined nearly 30\% from the July 2020 average as wildfires burned across the state in the first two weeks of September  2020. This trend is consistent with the elevated PM$_{2.5}$ level over the same time period as shown in the bottom row of Figure \ref{fig:motivation}.

\vspace{-20pt}

\begin{figure}[h!]
    \centering
    \includegraphics[width=0.8\textwidth]{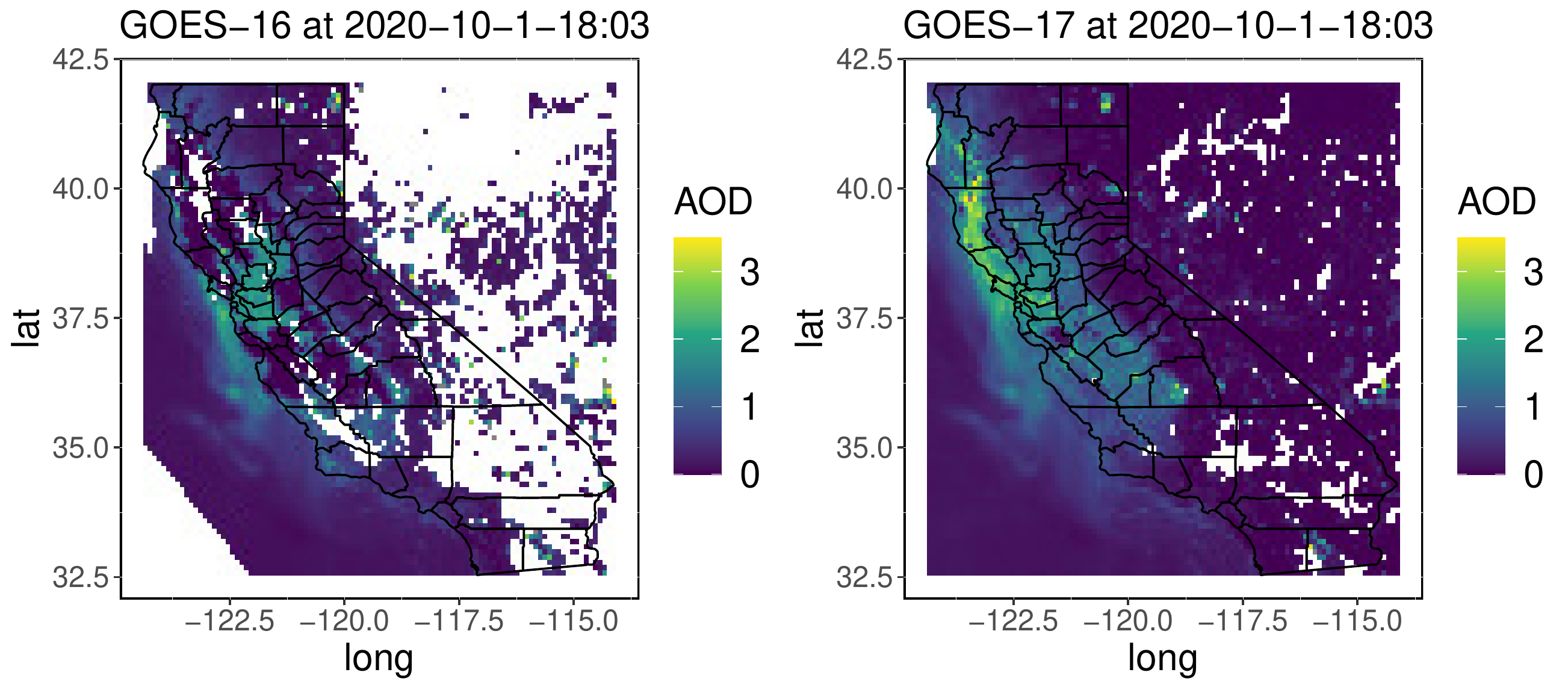}
    \caption{AOD images (taken at the same time over the same spatial area) from GOES-16 and GOES-17 geostationary satellites, where the heterogeneity between the two images is clearly visible (e.g., data missing rates and AOD readings)}
    \label{fig:CAaod1617}
\end{figure}

Atmospheric aerosols are measured by Aerosol Optical Depth (AOD)---the aerosols distributed within a column of air from the Earth's surface to the top of the atmosphere. AOD measurements data streams can be retrieved from multiple geostationary remote-sensing satellites, enabling the real-time modeling and short-term prediction of AOD processes. For example, the Advanced Baseline Imager (ABI) on board the NOAA's GOES-16 and GOES-17 satellites both generate remote-sensing AOD images for the continental U.S. (CONUS) every 5 minutes. Figure \ref{fig:CAaod1617} shows the observed AOD from both GOES-16 and GOES-17 at 2020-10-01-18:03 over the West Coast. It is critical to note that, \textit{although the two images are taken at the same time over the same spatial area for the same underlying process, the two images present very heterogeneous characteristics}. For example, 

$\bullet$ The data missing rates are clearly different from the two data streams (the white spaces correspond to the areas over which no measurements are available). 

$\bullet$ Even the measured AOD values can be different. In the Northwest portions of these two images, GOES-16 (on the left) clearly shows much lower AOD readings than GOES-17 (on the right). In this example, GOES-16 fails to see the wildfire burning in the Northwest region of the image. 

\vspace{-8pt}

In fact, heterogeneity among multi-source data streams is a very common issue in optical remote-sensing  (e.g., data missing rate, sampling frequencies, systematic bias, measurement errors, etc.). Remote-sensing data streams are noisy and potentially biased observations of an underlying physical process. There exist pressing needs and theoretical interests to simultaneously utilize (or, fuse) multi-source remote-sensing data streams, so that the underlying aerosol propagation process can be accurately estimated and predicted. 

From the perspective of atmospheric science, the propagation of aerosols is an advection-diffusion process driven by wind. Capturing the advection-diffusion dynamics becomes indispensable for accurate and interpretable predictions. Understanding the interactions between aerosols and wind is also critical for determining the space-time correlation structure of the statistical model for aerosol propagation. This application example motivates us to investigate the physics-informed statistical modeling for multi-source spatial-temporal data (arising from the same underlying physical process), i.e., the modeling of wildfire aerosols propagation by utilizing multi-source geostationary satellite remote-sensing data streams. 

\subsection{Literature Review and Contributions}
Statistical approaches have been widely applied to model environmental monitoring data. For example, \citet{Kuusela} proposed a locally stationary Gaussian Process (GP) regression model for Argo float data in a moving-window fashion, which avoids explicitly modeling the non-stationary covariance structure. \citet{Guinness} proposed a statistical model that accounts for both spatial-temporal jumps and non-stationarity in dynamic temperature fields with high temporal resolution. \citet{Stroud2017} developed a Monte Carlo EM algorithm for GP that handles the missing data issue which is commonly encountered in environmental monitoring. For non-stationary and highly dynamic processes, \citet{Stroud2001} proposed a Gaussian State Space Model (GSSM) that allows regression parameters to evolve over time, such that non-stationary spatio-temporal data can be handled. Based on a scalar transport equation, \cite{Liu2016} proposed a spatio-temporal model for ground-level ozone concentration using a convolution approach.  \citet{Liu2018b} investigated the modeling of weather radar reflectivity data, in which a space-time autoregressive process is used to characterize the motion of storms and a spatio-temporal conditional autoregressive process is employed to describe the small-scale random variation. Interested readers may refer to \cite{Stein, Banerjee, Cressie} for a more comprehensive review of statistical spatio-temporal models.

In particular, for the statistical modeling of satellite remote-sensing data, \citet{Kang} developed a Bayesian spatial random-effects model for AOD data from the MISR instrument on NASA’s Terra satellite. \citet{Katzfuss2011} adopted a spatio-temporal mixed-effects model for CO$_2$ measurements from NASA’s Aqua satellite, in which a fixed-rank smoothing and EM algorithm are developed for parameter estimation. \cite{Ma2020a} combined the Gaussian Markov random field and fixed-rank kriging as a fused GP model, and applied this model to sea surface temperature measurements from the MODIS satellite data. Note that, when an environmental process is monitored by various instruments or satellites, data assimilation or fusion is often needed to fully utilize data from multiple sources. \citet{Nguyen2014} proposed a data fusion approach for CO$_2$ observations from Japan's Greenhouse gases Observing SATellite (GOSAT) and NASA's Atmospheric InfraRed Sounder (AIRS). Based on a spatio-temporal random-effect model, \cite{Ma2020b} developed a Dynamic Fused Gaussian Process (DFGP) for sea surface temperature data from MODIS and Advanced Microwave Scanning Radiometer-Earth Observing System (AMSR-E) instruments.

\vspace{8pt}
This paper further advances the state-of-the-art by making the following contributions.

$\bullet$ The proposed approach enables the simultaneous modeling of heterogenous multi-source data streams with different spatial resolutions, data quality, sampling frequencies, etc.; 

$\bullet$ The approach further improves the modeling accuracy and interpretability over a recently proposed physics-informed spatio-temporal model \citep{Liu2021} by addressing the inevitable bias due to the inadequacy of first-order physics and the truncation error of the Fourier series; 

$\bullet$ The paper successfully applies the proposed approach to a real data set, demonstrates the advantages of physics-informed statistical learning for a critical application, and makes the computer code readily available. 

\vspace{-8pt}
\begin{figure}[h!]
	\begin{center}
		\includegraphics[width=1.05\textwidth]{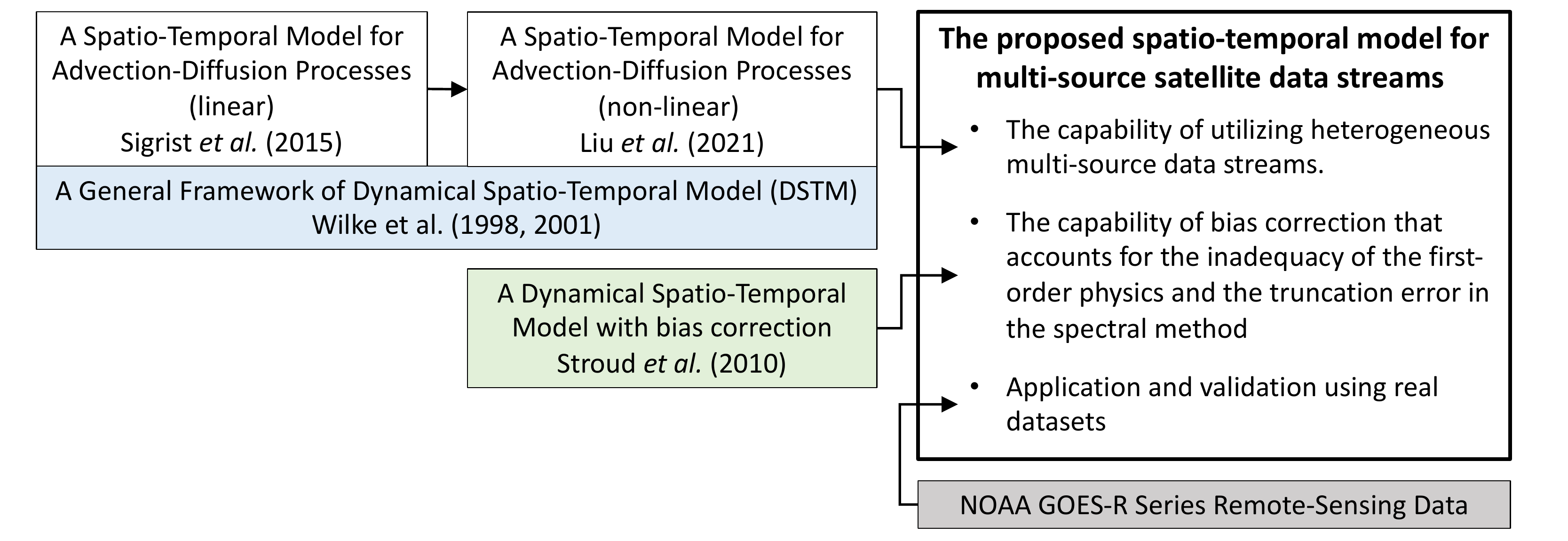}
	\end{center}
	\caption{Contributions of the current paper and its relationship with existing literature}
	\label{fig:literature} 
\end{figure}

\vspace{-8pt}
The connection between this work and the existing literature is sketched in Figure \ref{fig:literature}. Detailed discussions are provided as follows:

\cite{Sigrist} proposed a physics-based spatio-temporal model for advection-diffusion processes governed by a stochastic Partial Differential Equation (sPDE). The advection-diffusion operator of the sPDE does not change in space or time, resulting in a linear problem. Recently, \cite{Liu2021} extended the previous work by allowing the advection-diffusion operator to vary in space (i.e., the non-linear case). The model in \cite{Liu2021} is based on the spectral decomposition and can be conceptually written as:
\begin{align}\label{eq:liu2021}
	\begin{split}
		\bm{y}_t &= \bm{F}\bm{\theta}_t+\text{noise} \\
		\bm{\theta}_t &= \bm{G}\bm{\theta}_{t-1}+\text{noise},
	\end{split}
\end{align}
where $\bm{y}_t$ is a vector that contains observations at time $t$, $\bm{F}$ is a matrix of Fourier basis functions, $\bm{\theta}_t$ is the state vector that contains the temporal Fourier coefficients, and $\bm{G}$ is the transition matrix that is determined by governing physics. Note that, the Fourier coefficients in $\bm{\theta}_t$ are truncated, and only a limited number of low-frequency terms are retained. Also note that, in the non-linear case, the temporal evolution of spectral coefficients are all coupled. 
Both papers leverage the general Dynamic Spatio-Temporal Model (DSTM), which decouples the modeling of underlying processes and observation data \citep{Wikle1998, Katzfuss2020}. 
The hierarchical structure of DSTM naturally enables one to link heterogeneous multi-source remote-sensing AOD data to an underlying physical process, and effectively characterize the uncertainty associated with multiple data streams \citep{Wikle2001}.

In this paper, we provide a more holistic view that unifies the results in \cite{Sigrist} and \cite{Liu2021} based on the spectral theory; see subsection \ref{sec:model}. 
In addition, model (\ref{eq:liu2021}) assumes that the governing advection-diffusion equation adequately describes the underlying physical process, which is rarely true in reality. In fact, (first-order) governing equations are often approximations to complicated real-world processes, and bias is almost always present. For example, \cite{Stroud2010} introduced a bias correction term when investigating the Michigan lake sediment transport using a finite differences method. 
It is also noted that, since model (\ref{eq:liu2021}) approximates the observation using a spectral method that discards the high-frequency terms, bias is inevitably introduced into the model (e.g., the Gibbs phenomenon due to the discontinuity at the boundaries of the images \citep{Gibbs}). We extend model (\ref{eq:liu2021}) in \cite{Liu2021} by including a data-driven bias correction process when modeling the remote-sensing AOD data. 

\vspace{16pt}
The remainder of this paper is organized as follows. Section 2 presents the physics-informed statistical model for multi-source data streams. Section 3 presents an application example of the proposed model using GOES-16 and GOES-17 AOD image streams from the 2020 California Glass Fire. Comprehensive comparisons and discussions on the computational aspects of the proposed approach are presented. Section 4 concludes this paper. 

\vspace{8pt}
\section{A Physics-Informed Statistical Model for Multi-Source Remote-Sensing Data Streams}
\vspace{8pt}
\subsection{Basic Setup and Notation}
\begin{figure}[h!]
	\centering
	\includegraphics[width=0.8\textwidth]{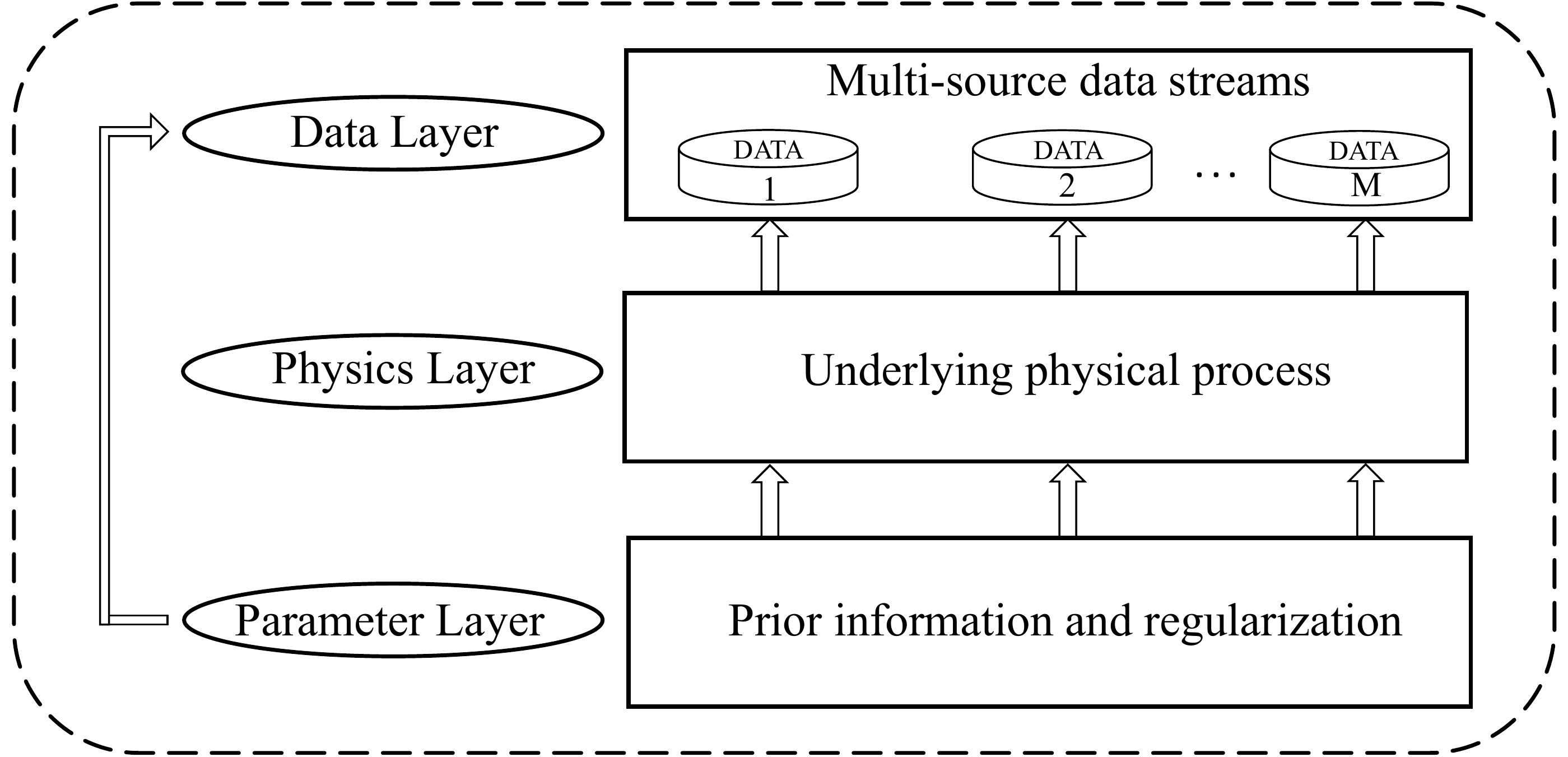}
	\caption{A hierarchical modeling structure for multi-source data streams arising from a physical process}\label{fig:Framework}
\end{figure}

As sketched in Figure \ref{fig:Framework}, the proposed physics-informed statistical model consists of three layers: a Physics Layer, a Data Layer and a Parameter Layer.

The Physics Layer captures the underlying physical aerosols propagation process. This layer, to be constructed from the governing physics, describes the dynamics of  hidden physical states of the aerosols propagation process: $[\bm{\theta}_t|\bm{\theta}_{t-1}, \bm{\psi}_P]$, where $\bm{\theta}_t$ is the state vector that characterizes the physical process (to be discussed later), $\bm{\psi}_P$ contains the unknown parameters, and $[\cdot]$ denotes the probability distribution.

The Data Layer models the observed multi-source remote-sensing data streams arising from the underlying
aerosols propagation process. In this paper, $M$ denotes the number of data sources. For each data source $m\in\{1,2,\cdots,M\}$, observations are available at $T^{(m)}=\{t^{(m)}_1,\dots,t^{(m)}_{T_m}\}$, where $T_m$ is the total number of temporal observations from source $m$. At any time $t\in T^{(m)}$, $\bm{y}^{(m)}_t$ is a $(N-n_t^{(m)}) \times 1$ vector representing the observations from $N$ spatial locations, where $n_t^{(m)}$ is the number of missing data points from source $m$ at that time. Let $\mathcal{T}=\bigcup_{j=1}^M T^{(m)}$ be the joint observation times from all data sources. Given any time $t\in\mathcal{T}$, let $J_t = \{m:t \cap T^{(m)}\neq \varnothing, m=1,2,\cdots,M\}$ be a collection of data sources where observations are available at that time. Then, all observations available at time $t$ are represented by concatenating the observations from individual sources, i.e., $\bm{y}_t=\text{vec}(\{ \bm{y}^{(m)}_t\}_{m \in J_t})$. 
Hence, the Data Layer is expressed by $[\bm{y}_t|\bm{\theta}_t, \bm{\psi}_D]$, where $\bm{\psi}_D$ contains the unknown parameters in this layer, and the observations depend on the state vector $\bm{\theta}_t$ of the underlying physical process. 
Note that, (\textit{i}) the separation of the Data Layer and the Physics Layer enables the handling of multi-modal or multi-source observations arising from the aerosols propagation process \citep{Wikle1998, Katzfuss2020}; (\textit{ii}) missing or non-gridded data can be handled by defining a mapping between the regular and irregular support (which is to be fully exploited in this paper); (\textit{iii}) uncertainties associated with multi-source data streams can be characterized by the observational errors.

The Parameter Layer enables the incorporation of additional regularizations or Bayesian priors on the parameters, $[\bm{\psi}_D, \bm{\psi}_P]$.

The following sub-sections present the technical details of how each layer can be modeled and integrated into a dynamic model for multi-source remote-sensing data streams. 

\subsection{The Model} \label{sec:model}
The propagation of wildfire aerosols is an advection-diffusion process described by a Partial Differential Equation (PDE) \citep{Cushman}:
\begin{equation}\label{eq:PDE} 
\dot{\xi}(\bm{s},t) = \mathcal{A}\xi(\bm{s},t)+ Q(\bm{s}, t), 
\end{equation} 
where $\xi(\bm{s},t)$ represents the AOD (Aerosol Optical Depth) process in space and time, $Q(\bm{s},t)$ is the source term (aerosol emission), and the advection-diffusion operator $\mathcal{A}$ is given by:
\begin{equation}\label{eq:operator} 
	\mathcal{A}\xi(\bm{s},t) = -\bm{v}^T(\bm{s},t)\nabla \xi(\bm{s},t)+\nabla\cdot[\bm{D}(\bm{s},t)\nabla \xi(\bm{s},t)]
\end{equation} 
with $\bm{v}(\bm{s},t)$, $\bm{D}(\bm{s},t)$, $\nabla$ and $\nabla\cdot$ respectively being the wind field, diffusivity, gradient and divergence. The PDE (\ref{eq:PDE}) has been wildly used for modeling the particle movement inside a physical system such as the sediment concentration in water \citep{Stroud2010} and pollutant dispersion in air \citep{Liu2016}. 
The first term on the right-hand side of  (\ref{eq:operator}) captures the advection of the wildfire aerosols propagation driven by a wind field $\bm{v}(\bm{s},t)$, while the second term of  (\ref{eq:operator}) represents the diffusion process driven by the concentration gradient. In this paper, both $\bm{v}(\bm{s},t)$ and $\bm{D}(\bm{s},t)$ may vary in space and time. Note that, although \citet{Liu2021} only considers spatially-varying but temporally-invariant advection-diffusion, we note that the framework can be extended to spatially- and temporally-varying advection-diffusion. The latter is more flexible and better captures the complex spatial and temporal dynamics of the short-term wildfire aerosols propagation.

\vspace{8pt}
We first provide a different perspective to the main results obtained from \cite{Liu2021}. Following the spectral theory, the eigenfunction solution of the PDE (\ref{eq:PDE}) is given by:
\begin{equation} \label{eq:eigenfunction}
	\xi(\bm{s},t) = \sum \alpha_{\bm{k}}(t)f_{\bm{k}}(\bm{s}), 
\end{equation}
where $f_{\bm{k}}(\bm{s})$ is the eigenfunction (e.g., the Fourier bases) and $ \alpha_{\bm{k}}(t)$ determines the weighting of each mode. Similarly, the source term $Q(\bm{s},t)$ can also be expressed by $Q(\bm{s},t) = \sum \beta_{\bm{k}}(t)f_{\bm{k}}(\bm{s})$.

$\diamond$ In the linear case (i.e., when the operator $\mathcal{A}$ does not change in space or time), taking the inner product of $\dot{\xi}(\bm{s},t)$ and the adjoint eigenfunction $\tilde{f}_{\bm{k}'}(\bm{s})$ yields (see page 43 of \citet{Kutz} for details):
\begin{equation} \label{eq:linear}
	\begin{split}
	\left\langle \dot{\xi}(\bm{s},t), \tilde{f}_{\bm{k}'}(\bm{s})\right\rangle  & = \left\langle \sum \dot{\alpha}_{\bm{k}}(t)f_{\bm{k}}(\bm{s}) , \tilde{f}_{\bm{k}'}(\bm{s})\right\rangle  \\
	& = \lambda_{\bm{k}'}^*\left\langle \sum \alpha_{\bm{k}}(t)f_{\bm{k}}(\bm{s}) , \tilde{f}_{\bm{k}'}(\bm{s})\right\rangle + \left\langle\sum {\beta}_{\bm{k}}(t)f_{\bm{k}}(\bm{s}) , \tilde{f}_{\bm{k}'}(\bm{s})\right\rangle ,
	\end{split}
\end{equation}
which implies that $\dot{\alpha}_{\bm{k}}(t) = \lambda_{\bm{k}'}^* \alpha_{\bm{k}}(t) + {\beta}_{\bm{k}}(t)$, where $\lambda_{\bm{k}'}^*$ is the conjugate adjoint eigenvalues. Hence, for any $\bm{k}$, the temporal evolution of $\alpha_{\bm{k}}(t)$ only depends on itself. This is consistent with the main result presented in \cite{Sigrist}. 

$\diamond$ In the non-linear case (i.e., when the operator $\mathcal{A}$ varies in space and time),  taking the inner product of $\dot{\xi}(\bm{s},t)$ and the adjoint eigenfunction $\tilde{f}_{\bm{k}'}(\bm{s})$ yields
\begin{equation}
		\left\langle \dot{\xi}(\bm{s},t), \tilde{f}_{\bm{k}'}(\bm{s})\right\rangle   = 	\left\langle \sum \dot{\alpha}_{\bm{k}}(t)f_{\bm{k}}(\bm{s}) , \tilde{f}_{\bm{k}'}(\bm{s})\right\rangle  = \left\langle \mathcal{A}\xi(\bm{s},t), \tilde{f}_{\bm{k}'}(\bm{s})\right\rangle.
\end{equation}
Because $\mathcal{A}$ is non-linear, and we can no longer obtain the second row of (\ref{eq:linear}). Hence, for any $\bm{k}$, the temporal evolution of  $\alpha_{\bm{k}}(t)$ depends on all spectral coefficients in (\ref{eq:eigenfunction}). \cite{Liu2021} provided the closed-form expression of the transition matrix that determines the dynamics of $\alpha_{\bm{k}}(t)$. 

\vspace{8pt}
In this paper, we consider a real-valued process $\xi(\bm{s},t)$ defined on a $N_1\times N_2$ uniform spatial grid at discrete times (without loss of generality, $N_1$ and $N_2$ are assumed to be even numbers). The spatial domain is defined by a Cartesian product $\mathbb{S} = A\times B = \{\bm{s}=(s_1, s_2)|s_1\in A, s_2\in B\}$, where $A = \{0,1/N_1,\dots, (N_1-1)/N_1\}$ and $B=\{0, 1/N_2,\dots,(N_2-1)/N_2\}$. The spectral method decomposes the process $\xi(\bm{s},t)$ into a linear combination of orthogonal spatial basis functions and temporal coefficients that evolve over time \citep{Liu2021}:
\begin{equation} \label{eq:decomposition2}
	\xi(\bm{s},t) = \sum_{\bm{k}\in\mathcal{K}_1}\alpha_{\bm{k}}^{(c)}(t)f_{\bm{k}}^{(c)}(\bm{s}) + 2\sum_{\bm{k}\in\mathcal{K}_2} \left(\alpha_{\bm{k}}^{(c)}(t)f_{\bm{k}}^{(c)}(\bm{s}) + \alpha_{\bm{k}}^{(s)}(t)f_{\bm{k}}^{(s)}(\bm{s})\right),
\end{equation}
where $f_{\bm{k}}^{(c)}(\bm{s})=\cos(\bm{k}^T\bm{s})$ and $f_{\bm{k}}^{(s)}(\bm{s})=\sin(\bm{k}^T\bm{s})$ are the Fourier basis functions for real-valued processes, $\alpha_{\bm{k}}^{(c)}(t)$ and $\alpha_{\bm{k}}^{(s)}(t)$ are the coefficients respectively corresponding to $f_{\bm{k}}^{(c)}(\bm{s})$ and $f_{\bm{k}}^{(s)}(\bm{s})$, $\bm{k}$ is the spatial wave number, and $\mathcal{K}_1$ and $\mathcal{K}_2$ are two disjoint subsets:  
\begin{equation}\label{eq:Omega12}
	\begin{split}
	\mathcal{K}_1 &= \left\{(0,0), \left(0, N_2\big/2\right), \left(N_1\big/2, 0\right), \left(N_1\big/2, N_2\big/2\right)\right\}\nonumber\\
\mathcal{K}_2 &= \left\{(k_1,k_2)|k_1=1,\cdots, N_1\big/2-1, k_2=-N_2\big/2+1,\cdots, -1\right\}\nonumber\\
	&\hskip0.2in\cup\left\{(k_1,k_2)|k_1=0,1,\cdots, N_1\big/2, k_2=1,2,\cdots, N_2\big/2-1\right\}\nonumber\\
	&\hskip0.2in\cup\left\{\left(k_1,N_2\big/2\right)|k_1=0,1,\cdots, N_1\big/2\right\},
	\end{split}
\end{equation}
where $|\mathcal{K}_1|+2|\mathcal{K}_2|=N_1\times N_2$.

The source term $Q(\bm{s}, t)$ is described by a linear model $Q(\bm{s}, t)=z(\bm{s})b(t)$, where $z(\bm{s})$ is a spatial covariate and $b(t)$ is the coefficient. If the covariate process admits a spectral representation 
\begin{equation}
    z(\bm{s})  = \sum_{\bm{k}\in\mathcal{K}_1}\beta_{\bm{k}}^{(c)}f_{\bm{k}}^{(c)}(\bm{s}) + 2\sum_{\bm{k}\in\mathcal{K}_2} \left(\beta_{\bm{k}}^{(c)}f_{\bm{k}}^{(c)}(\bm{s}) + \beta_{\bm{k}}^{(s)}f_{\bm{k}}^{(s)}(\bm{s})\right),
\end{equation}
the linear model $Q(s, t) = z(s)b(t)$ implies
\begin{equation}\label{eq:source}
	Q(\bm{s}, t)  = \sum_{\bm{k}\in\mathcal{K}_1}\tilde{\beta}_{\bm{k}}^{(c)}(t)f_{\bm{k}}^{(c)}(\bm{s}) + 2\sum_{\bm{k}\in\mathcal{K}_2} \left(\tilde{\beta}_{\bm{k}}^{(c)}(t)f_{\bm{k}}^{(c)}(\bm{s}) + \tilde{\beta}_{\bm{k}}^{(s)}(t)f_{\bm{k}}^{(s)}(\bm{s})\right),
\end{equation}
where $\tilde{\beta}_{\bm{k}}^{(c)}(t)=\beta_{\bm{k}}^{(c)}b(t)$ and $\tilde{\beta}_{\bm{k}}^{(s)}(t)=\beta_{\bm{k}}^{(s)}b(t)$. In this paper, it is assumed that $b(t)$ is a Brownian motion with a constant mean. 

\vspace{8pt}
Let $\bm{k}_{1,i}$ be the $i$-th component in set $\mathcal{K}_1$, and $\bm{k}_{2,i}$ be the $i$-th component in set $\mathcal{K}_2$. We define $\bm{\alpha}_t=\text{vec}(\bm{\alpha}^{(c)}_{\mathcal{K}_1}(t), \bm{\alpha}^{(c)}_{\mathcal{K}_2}(t), \bm{\alpha}^{(s)}_{\mathcal{K}_2}(t))$, where 
$\bm{\alpha}^{(c)}_{\mathcal{K}_1}(t)=(\alpha^{(c)}_{\bm{k}_{1,1}}(t), \alpha^{(c)}_{\bm{k}_{1,2}}(t),\cdots, \alpha^{(c)}_{\bm{k}_{1,{|\mathcal{K}_1|}}}(t))^T$,
$\bm{\alpha}^{(c)}_{\mathcal{K}_2}(t)=(\alpha^{(c)}_{\bm{k}_{2,1}}(t), \alpha^{(c)}_{\bm{k}_{2,2}}(t),\cdots, \alpha^{(c)}_{\bm{k}_{2,{|\mathcal{K}_2|}}}(t))^T$, and 
$\bm{\alpha}^{(s)}_{\mathcal{K}_2}(t)=(\alpha^{(s)}_{\bm{k}_{2,1}}(t), \alpha^{(s)}_{\bm{k}_{2,2}}(t),\cdots, \alpha^{(s)}_{\bm{k}_{2,{|\mathcal{K}_2|}}}(t))^T$.
Similarly, we let $\tilde{\bm{\beta}}_t=\text{vec}(\tilde{\bm{\beta}}^{(c)}_{\mathcal{K}_1}(t), \tilde{\bm{\beta}}^{(c)}_{\mathcal{K}_2}(t), \tilde{\bm{\beta}}^{(s)}_{\mathcal{K}_2}(t))$, where 
$\tilde{\bm{\beta}}^{(c)}_{\mathcal{K}_1}(t)=(\tilde{{\beta}}^{(c)}_{\bm{k}_{1,1}}(t), \tilde{{\beta}}^{(c)}_{\bm{k}_{1,2}}(t), \cdots, \tilde{{\beta}}^{(c)}_{\bm{k}_{1,{|\mathcal{K}_1|}}}(t))^T$,
$\tilde{\bm{\beta}}^{(c)}_{\mathcal{K}_2}(t)=(\tilde{{\beta}}^{(c)}_{\bm{k}_{2,1}}(t), \tilde{{\beta}}^{(c)}_{\bm{k}_{2,2}}(t), \cdots, \tilde{\beta}^{(c)}_{\bm{k}_{2,{|\mathcal{K}_2|}}}(t))^T$, and 
$\tilde{\bm{\beta}}^{(s)}_{\mathcal{K}_2}(t)=(\tilde{\beta}^{(s)}_{\bm{k}_{2,1}}(t),\tilde{\beta}^{(s)}_{\bm{k}_{2,2}}(t),\cdots, \tilde{\beta}^{(s)}_{\bm{k}_{2,{|\mathcal{K}_2|}}}(t))^T$. Then, after projecting the PDE (\ref{eq:PDE}) to the Fourier bases and using the Galerkin method, the dynamics of the temporal coefficients can be captured by an Ordinary Differential Equation (ODE):
\begin{equation}\label{eq:temporalcoeffODE}
	\bm{\dot{\alpha}}_t = \bm{P}_t\bm{\alpha}_t + \tilde{\bm{\beta}}_t,
\end{equation}
where the transition matrix $\bm{P}_t$ is given by
\begin{equation}\label{eq:P}
\bm{P}_t = \begin{pmatrix}
	\bm{C}_1^{-1}\bm{\Psi}^{(\text{A1+D1})}_{11}& 2\bm{C}_{1}^{-1}\bm{\Psi}^{(\text{A1+D1})}_{12}& 2\bm{C}_{1}^{-1}\bm{\Psi}^{(\text{A2+D2})}_{12}\\
	\frac{1}{2}\bm{C}_{2}^{-1}\bm{\Psi}^{(\text{A1+D1})}_{21}& 
	\bm{C}_{2}^{-1}\bm{\Psi}^{(\text{A1+D1})}_{22}&
	\bm{C}_{2}^{-1}\bm{\Psi}^{(\text{A2+D2})}_{22}\\
	\frac{1}{2}\bm{C}_{2}^{-1}\bm{\Psi}^{(\text{A3+D3})}_{21}& 
	\bm{C}_{2}^{-1}\bm{\Psi}^{(\text{A3+D3})}_{22}& 
	\bm{C}_{2}^{-1}\bm{\Psi}^{(\text{A4+D4})}_{22}
\end{pmatrix}.
\end{equation}

Note that, the transition matrix $\bm{P}_t$ is completely determined by the physics process (\ref{eq:PDE}) and depends on the advection and diffusion parameters, $\bm{v}(\bm{s},t)$ and $\bm{D}(\bm{s},t)$. Detailed expressions of $\bm{P}_t$ are provided in Appendix A. 

\vspace{8pt}
Although (\ref{eq:temporalcoeffODE}) captures the dynamics of $\bm{\alpha}_t$ based on the physics process (\ref{eq:PDE}), it is worth noting that systematic bias may exist in (\ref{eq:temporalcoeffODE}) due to the following two major considerations. 

\begin{figure}[ht]
	\centering
	\includegraphics[width=1\textwidth]{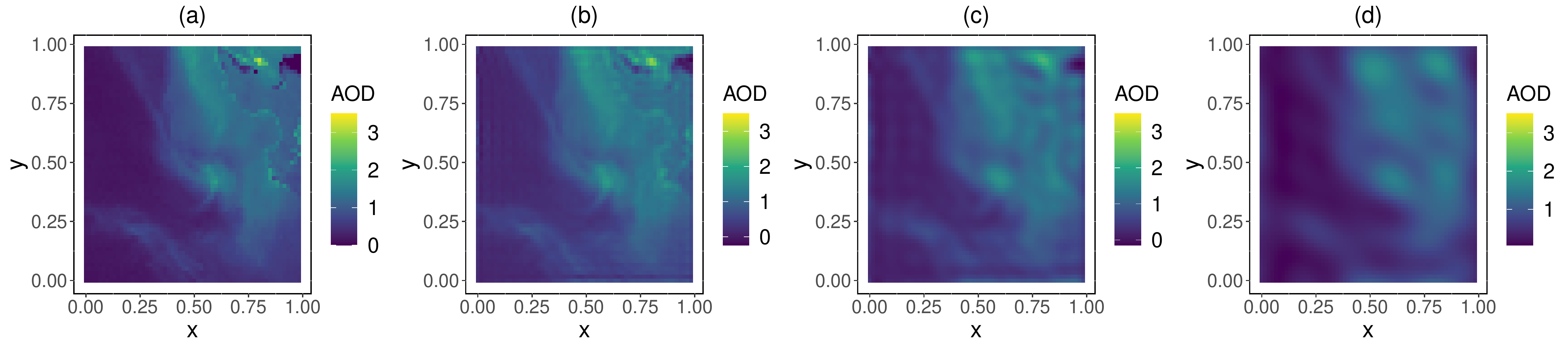}
	\caption{Artifacts due to the truncation of high-frequency Fourier terms: (a) original AOD image; (b)-(d) reconstructed  images using IFT with the wavenumbers being truncated at $(40, 40), (20, 20), (10, 10)$ respectively.}\label{fig:IFT}
\end{figure}

$\bullet$ The transition matrix $\bm{P}_t$ in (\ref{eq:P}) is a dense matrix of a dimension of  $N_1\times N_2$. As shown in Appendix A, the evaluation of each element in $\bm{P}_t$ requires double numerical integration over the spatial domain and is computationally expensive. The conventional strategy is to truncate the wavenumbers by only keeping a (much smaller) number of low-frequency terms. However, dropping high-frequency Fourier terms inevitably leads to artifacts (i.e., the well-known Gibbs phenomenon when re-constructing the aerosols process through Inverse Fourier Transform (IFT) \citep{Gibbs}) and introduces errors into the transition equation. For illustrative purposes, Figure \ref{fig:IFT}(a) shows a satellite image with $60 \times 60$ pixels ($N_1=N_2=60$) from the GOES-17 AOD data product. Figure \ref{fig:IFT}(b)-(d) are the AOD images recovered by IFT with wavenumbers being truncated at (40, 40), (20, 20) and (10, 10), respectively. Figure \ref{fig:IFT}(b)-(d) clearly show that it becomes more difficult to capture the small-scale characteristics in the original AOD image if more high-frequency terms are discarded.

$\bullet$ The second source of systematic bias is due to the fact that first-order physics (such as the PDE (\ref{eq:PDE})) is only an approximation of the complex real-world process. In addition, the calculation of the transition matrix $\bm{P}_t$ depends on the velocity field $\bm{v}(\bm{s},t)$ and diffusivity $\bm{D}$, which are never precisely known and are often estimated from data. This means that even if no truncation is made (i.e., all Fourier coefficients are remained), systematic bias still presents in (\ref{eq:temporalcoeffODE}). 

\vspace{8pt}
Hence, a bias correction term $\bm{\gamma}_t\Delta$ is added to the discrete-time (first-order) solution of (\ref{eq:temporalcoeffODE}):
\begin{equation}
	\bm{\alpha}_{t+1} = \exp(\bm{P}_t\Delta)\bm{\alpha}_t+\tilde{\bm{\beta}}_t\Delta+\bm{\gamma}_t\Delta+\bm{w}_{t}^{(\alpha)},
\end{equation}
where $\text{exp}(\cdot)$ is the matrix exponential, $\bm{w}_{t}^{(\alpha)}$ is the random noise that captures the uncertainty associated with the transition of $\bm{\alpha}_{t}$, and $\Delta$ is the time interval. Without loss of generality, we let $\Delta=1$ in all subsequent discussions. 

Let $\bm{\theta}_t^\ast = (\bm{\alpha}_t^T, \bm{\tilde{\bm{\beta}}}_t^T, \bm{\gamma}_t^T)^T$ be the augmented state vector, we obtain:
\begin{equation}\label{eq:augmentedODE}
	\bm{\theta}_{t+1}^\ast = \bm{G}_t^\ast\bm{\theta}_t^\ast + \bm{w}_t^\ast,
\end{equation}
where 
$\bm{G}_t^\ast=
\begin{pmatrix}
	\exp(\bm{P}_t) & \bm{I} & \bm{I}\\
	\textbf{0} &  \bm{I} &\textbf{0}\\
	\textbf{0} & \textbf{0} & \bm{I}
\end{pmatrix}$, 
$\bm{w}_t^\ast=
\begin{pmatrix}
	\bm{w}_{t}^{(\alpha)}\\
	\bm{w}_{t}^{(\beta)}\\
	\bm{w}_{t}^{(\gamma)}
\end{pmatrix}$, 
and $\bm{w}^\ast_t\sim \mathcal{N}(\textbf{0},\bm{W}^\ast)$.

\vspace{8pt}
The first row of $\bm{G}_t^\ast$ shows that $\bm{\alpha}_{t+1}$ depends on $\bm{\alpha}_{t}$ and the sum of $\bm{\tilde{\beta}}_t$ and $\bm{\gamma}_t$. Because $\bm{\tilde{\beta}}_t$ and $\bm{\gamma}_t$ are  two independent Brownian processes with constant (unknown) means, $\bm{\tilde{\beta}}_t$ and $\bm{\gamma}_t$ are not identifiable without additional regularizations. 
Hence, we let $\tilde{\bm{\gamma}}_t = \bm{\tilde{\beta}}_t+ \bm{\gamma}_t$, and re-write (\ref{eq:augmentedODE}) as:
\begin{align}\label{eq:degeneratedODE}
	\bm{\theta}_{t+1} = \bm{G}_t\bm{\theta}_t + \bm{w}_t,
\end{align}
where $\bm{\theta}_t=
\begin{pmatrix}
	\bm{\alpha}_t\\
	\tilde{\bm{\gamma}}_t
\end{pmatrix}$,
$\bm{G}_t=
\begin{pmatrix}
	\exp(\bm{P}_t) & \bm{I}\\
	\textbf{0} & \bm{I}
\end{pmatrix}$,
$\bm{w}_t=
\begin{pmatrix}
	\bm{w}_{t}^{(\alpha)}\\
	\bm{w}_{t}^{(\tilde{\gamma})}
\end{pmatrix}$,
and $\bm{w}_t\sim \mathcal{N}(\textbf{0}, \bm{W})$.

\vspace{8pt}
Finally, by combining (\ref{eq:decomposition2}),  (\ref{eq:source}) and (\ref{eq:degeneratedODE}), we obtain a dynamic model for multi-source satellite data streams arising from the same underlying process:
\begin{equation}\label{eq:dynamicalmodel}
	\begin{split}
	\bm{y}_t &= \bm{F}_t\bm{\theta}_t + \bm{v}_t, \hskip0.5in \bm{v}_t\sim \mathcal{N}(\textbf{0},\bm{V}_t) \\
	\bm{\theta}_t  &= \bm{G}_t\bm{\theta}_{t-1} + \bm{w}_{t-1}, \hskip0.11in \bm{w}_{t}\sim \mathcal{N}(\textbf{0},\bm{W}_t),
	\end{split}
\end{equation}
where $\bm{V}_t = \text{diag}(\{\sigma_m^2\bm{I}\}_{m=1}^{|J_t|})$ and $\bm{W}_t$ are covariance matrices, and
\begin{equation}
	 \bm{F}_t = \begin{pmatrix}
	 	\bm{F}^{(1)}_t \\
	 	 \bm{F}^{(2)}_t \\
	 	 \vdots \\
	 	 \bm{F}^{(|J_t|)}_t 
	 \end{pmatrix} =
     \begin{pmatrix}
    	\bm{K}^{(1)}_t \bm{F}, \bm{0}\\
    	\bm{K}^{(2)}_t \bm{F}, \bm{0}\\
    	\vdots \\
    	\bm{K}^{(|J_t|)}_t \bm{F}, \bm{0}
    \end{pmatrix}.
\end{equation}

In addition, 

$\diamond$  $J_t$ is a collection of data sources where observations are available at time $t$.

$\diamond$ $\bm{F}=(\bm{f}(\bm{s}_1),\cdots,\bm{f}(\bm{s}_{N}))^T\in\mathbb{R}^{N\times (2|\mathcal{K}_1|+|\mathcal{K}_2|)}$, $\bm{f}(\bm{s})=(\bm{f}_{\mathcal{K}_1}^{(c)}(\bm{s}), 2\bm{f}_{\mathcal{K}_1}^{(c)}(\bm{s}), 2\bm{f}_{\mathcal{K}_2}^{(c)}(\bm{s}))^T$, where $\bm{f}_{\mathcal{K}_1}^{(c)}(\bm{s})=(f^{(c)}_{\bm{k}_{1,1}}(\bm{s}),f^{(c)}_{\bm{k}_{1,2}}(\bm{s})\cdots,f^{(c)}_{\bm{k}_{1,|\mathcal{K}_1|}}(\bm{s}))$, $\bm{f}_{\mathcal{K}_2}^{(c)}(\bm{s})=(f^{(c)}_{\bm{k}_{2,1}}(\bm{s}),f^{(c)}_{\bm{k}_{2,2}}(\bm{s}),\cdots,f^{(c)}_{\bm{k}_{2,|\mathcal{K}_2|}}(\bm{s}))$, and
$\bm{f}_{\mathcal{K}_2}^{(s)}(\bm{s})=(f^{(s)}_{\bm{k}_{2,1}}(\bm{s}),f^{(s)}_{\bm{k}_{2,2}}(\bm{s}),\cdots,f^{(s)}_{\bm{k}_{2,|\mathcal{K}_2|}}(\bm{s}))$.

$\diamond$ $\bm{K}^{(m)}_t$ is a $(N - n_t^{(m)}) \times N$ Boolean mapping matrix that handles the missing data issue for the available data source $m$ at time $t$ (i.e. $m\in J_t$), where $n_t^{(m)}$ is the number of missing data points at time $t$ from data source $m$.

\vspace{8pt}
The model (\ref{eq:dynamicalmodel}) contains the unknown parameters $\bm{\psi} = (\bm{\psi}_P, \bm{\psi}_D)$ and the state vector $\bm{\theta}_t$. Here, $\bm{\psi}_P$ is a collection of the 
entries of the dense covariance $\bm{W}$ and $\bm{\psi}_D=\{\sigma^2_{1}, \sigma^2_{2}, \cdots, \sigma^2_{M}\}$ is associated with $\bm{V}_t$. 

$\bullet$ Conditioning on $\bm{\psi}$,  the posterior distribution for $\bm{\theta}_t$ can be obtained using the Forward Filtering and Backward Sampling (FFBS) approach.  The FFBS algorithm employs the Kalman filter to get a one-step-ahead posterior of the state vector, and the backward sampling to obtain the full posterior given all observations $\bm{y}_{1:T}$. Let $\bm{m}_{t|t}$ and $\bm{C}_{t|t}$  denote the filtering mean and covariance matrix of $\bm{\theta}_t$ given the observations up to time $t$, and let  $\bm{m}_{t|t-1}$ and $\bm{C}_{t|t-1}$ be the one-step-ahead predictive mean and covariance matrix of $\bm{\theta}_t$ given the observations up to time $t-1$, we have
\begin{equation}
	\bm{m}_{t|t-1}  = \bm{G}_t\bm{m}_{t-1|t-1}, \quad\quad
	\bm{C}_{t|t-1}  = \bm{G}_t\bm{C}_{t-1|t-1}\bm{G}_t^T + \bm{W}
\end{equation}
with the initial condition $\bm{\theta}_0\sim \mathcal{N}(\bm{m}_{0|0},\bm{C}_{0|0})$. The filtering mean and covariance matrix given the observations up to time $t$ are 
\begin{equation}
    \begin{split}
	\bm{m}_{t|t}  &= \bm{m}_{t|t-1} + \bm{C}_{t|t-1}\bm{F}_t^T(\bm{F}_t\bm{C}_{t|t-1}\bm{F}^T_t+\bm{V}_t)^{-1}(\bm{y}_t-\bm{F}_t\bm{m}_{t|t-1})\\
	\bm{C}_{t|t}  &= \bm{C}_{t|t-1} - \bm{C}_{t|t-1}\bm{F}_t^T(\bm{F}_t\bm{C}_{t|t-1}\bm{F}_t^T+\bm{V}_t)^{-1}\bm{F}_t\bm{C}_{t|t-1}.
	\end{split}
\end{equation}

Then, the FFBS approach gives the posterior distribution of $\bm{\theta}_{1:T}$, i.e., $\bm{\theta}_{1:T}|\bm{y}_{1:T}$ as follows: 

$\diamond$ At time $T$, $\bm{\theta}_T|\bm{y}_{1:T} \sim \mathcal{N}(\bm{m}_{t|t}, \bm{C}_{t|t})$.

$\diamond$ For $t=T-1,T-2,\cdots,1$, $\bm{\theta}_t|\bm{\theta}_{t+1},\bm{y}_{1:T}\sim \mathcal{N}(\bm{h}_t, \bm{H}_t)$, where
\begin{equation}
    \begin{split}
	\bm{h}_t &= \bm{m}_{t|t} + \bm{C}_{t|t} \bm{G}^{T} \bm{C}_{t+1|t}^{-1}\left(\bm{\theta}_{t+1}-\bm{m}_{t+1|t}\right)\\
	\bm{H}_t &= \bm{C}_{t|t} - \bm{C}_{t|t}\bm{G}_{t+1}^T\bm{C}_{t+1|t}^{-1}\bm{G}_{t+1}\bm{C}_{t|t}.
	\end{split}
\end{equation}

$\bullet$ Conditioning on $\bm{\theta}$, we obtain (\textit{i}) the closed-form expression of the posterior distribution of $\bm{\psi}_P$ leveraging the conjugate inverse Wishart distribution for $\bm{W}$, i.e., $\bm{W}\sim \mathcal{W}^{-1}(\bm{\Phi}, \nu)$, and (\textit{ii}) the closed-form expression of the posterior distribution of $\bm{\psi}_D$ leveraging the conjugate inverse Gamma distribution, i.e., $\sigma^2_m \sim \mathcal{IG}\left(a_m,b_m\right)$ for $m=1,2,\cdots,M$.

The Gibbs sampler with FFBS for estimating the posterior distribution of $\bm{\psi} = (\bm{\psi}_P, \bm{\psi}_D)$ and $\bm{\theta}_t$ is summarized in Appendix B.

\section{Application: 2020 California Glass Fire}\label{sec:application}
\vspace{8pt}
\subsection{Background and Data Processing}
In this application example, we apply the proposed approach to model the California Glass Fire aerosols propagation using the GOES datasets. The Glass Fire started on September 27, 2020, and burned over 67,484 acres in the Wine Country, CA, for 23 days. This case study utilizes data from both GOES-16 and GOES-17, which were launched in November 2016 and March 2018, and collaboratively developed and operated by NASA and NOAA. The operational location of GOES-16 (GOES-17) is 75.2\textdegree W (137.2\textdegree W), and the main scanning area of GOES-16 (GOES-17) includes the CONUS (PACUS: Pacific Ocean including Hawaii), covering the state of California. 

\begin{figure}[ht]
    \centering
    \includegraphics[width=0.4\textwidth]{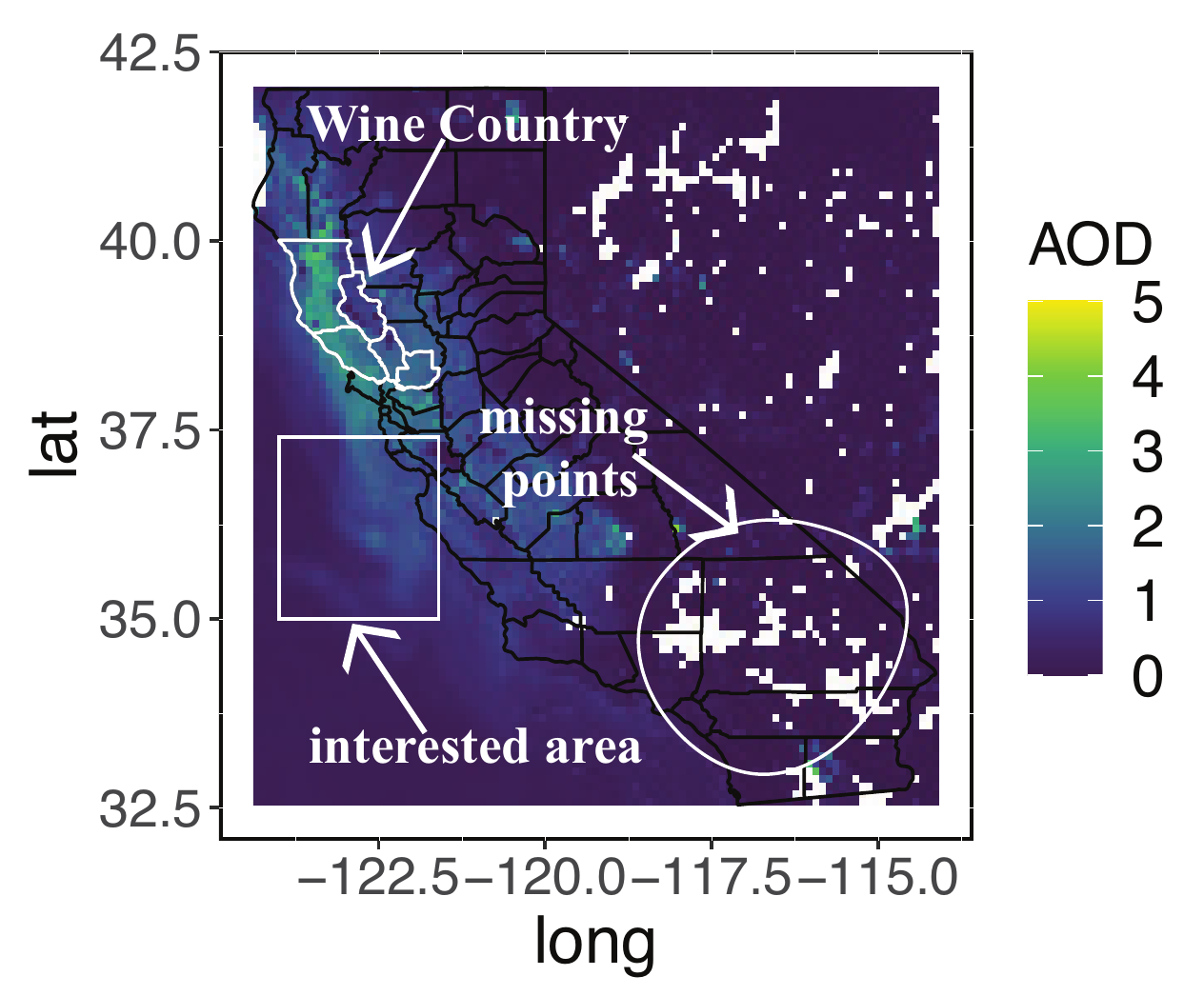}
    \vspace{-6pt}
    \caption{An AOD satellite image of the Glass fire at 18:56, Oct. 1, 2020.}\label{fig:Glass_fire}
\end{figure}

Figure \ref{fig:Glass_fire} shows an AOD satellite image from GOES-17 at 18:56, October 1, which is retrieved at 550nm channel (available on Google Cloud in the ABI-L2-AODC folder \citep{Google}). AOD is dimensionless and ranges from -0.05 to 5 in the GOES data product. An AOD value less than 0.1 indicates a crystal clear sky with the maximum visibility, and a value of 0.4 corresponds to a hazy condition \citep{Nguyen2012}. 
The AOD value over the Wine Country ranges from 3 to 5, indicating extremely haze conditions due to the wildfire. Note that, missing data (i.e., the white areas) are clearly observed in the image, which is a common issue for optical remote sensing.

\vspace{-12pt}

\vspace{-12pt}
\begin{figure}[ht]
	\centering
	\begin{subfigure}[t]{0.31\linewidth}
		\centering
		\subcaption{Pixel centers}
		\includegraphics[scale=0.42]{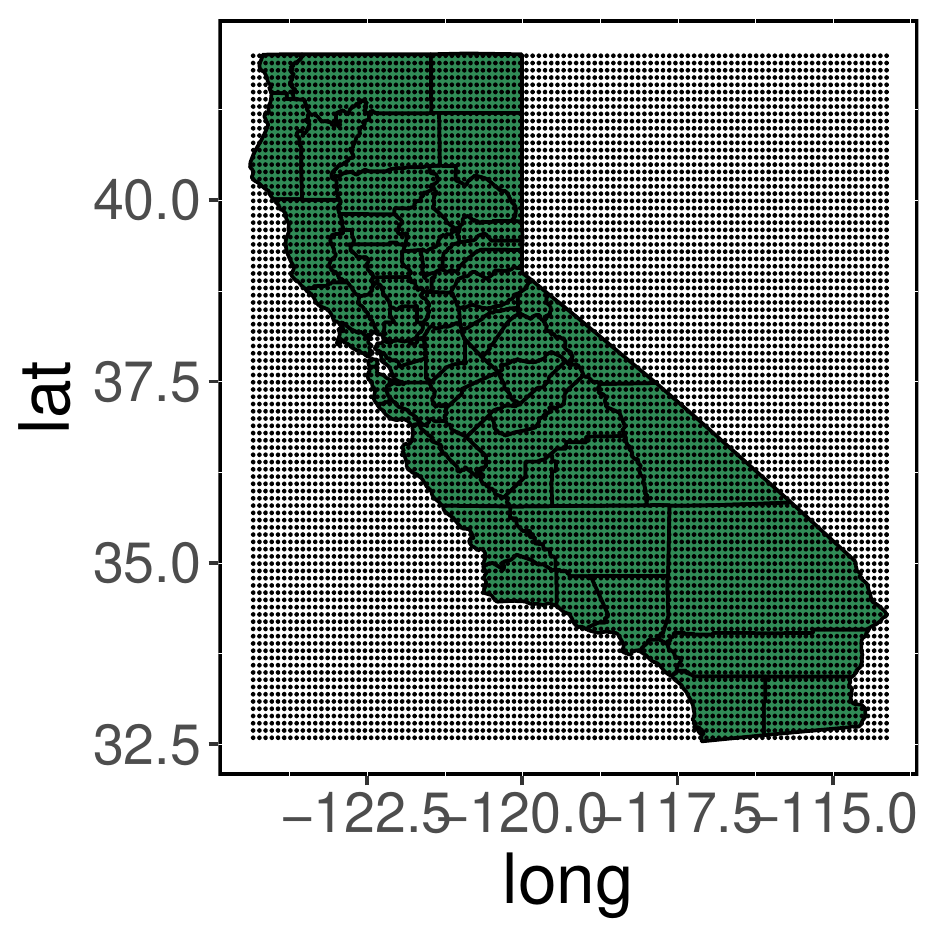}
	\end{subfigure}
	\quad
	\begin{subfigure}[t]{0.31\linewidth}
		\centering
		\subcaption{2020-10-01-18:03}
		\includegraphics[scale=0.42]{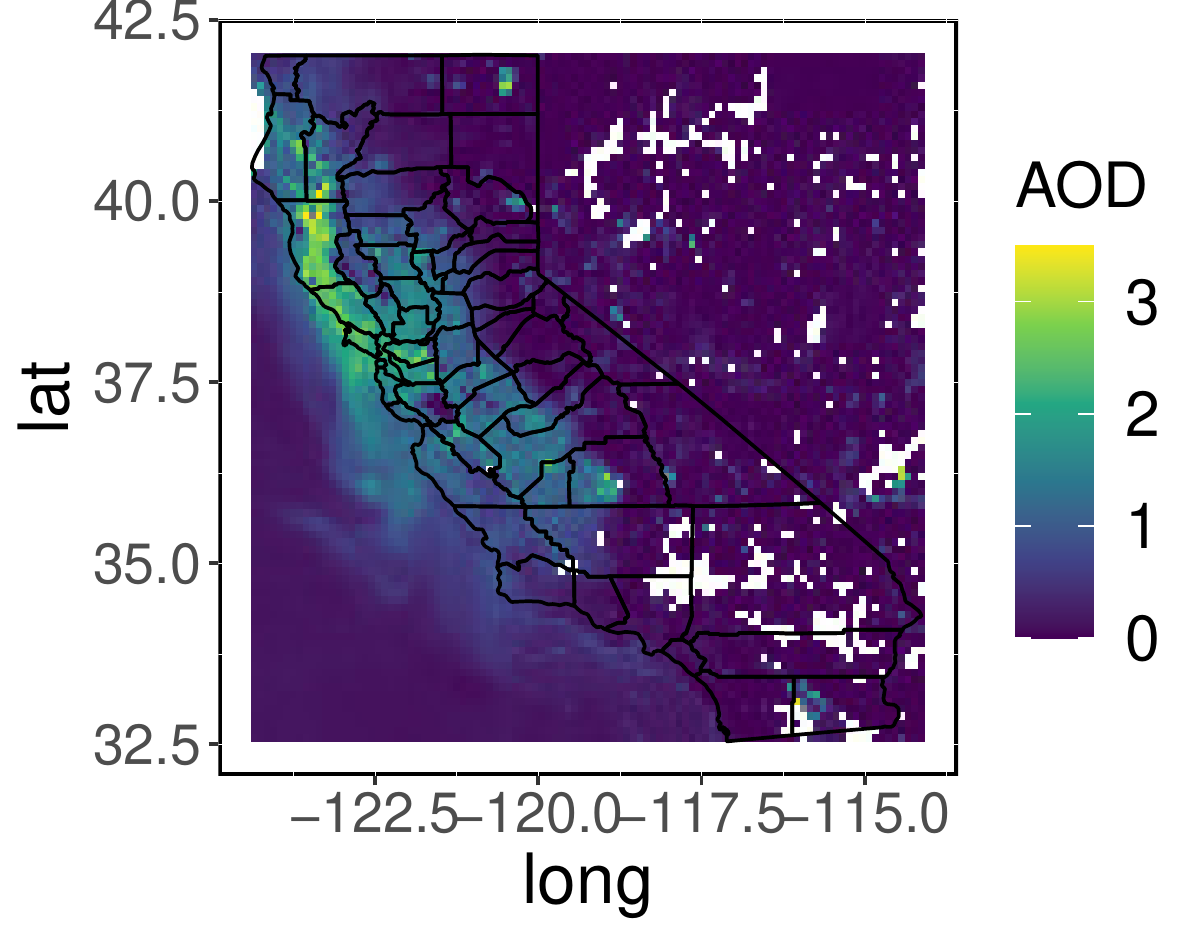}
	\end{subfigure}
	\quad
	\begin{subfigure}[t]{0.31\linewidth}
		\centering
		\subcaption{2020-10-01-18:08}
		\includegraphics[scale=0.42]{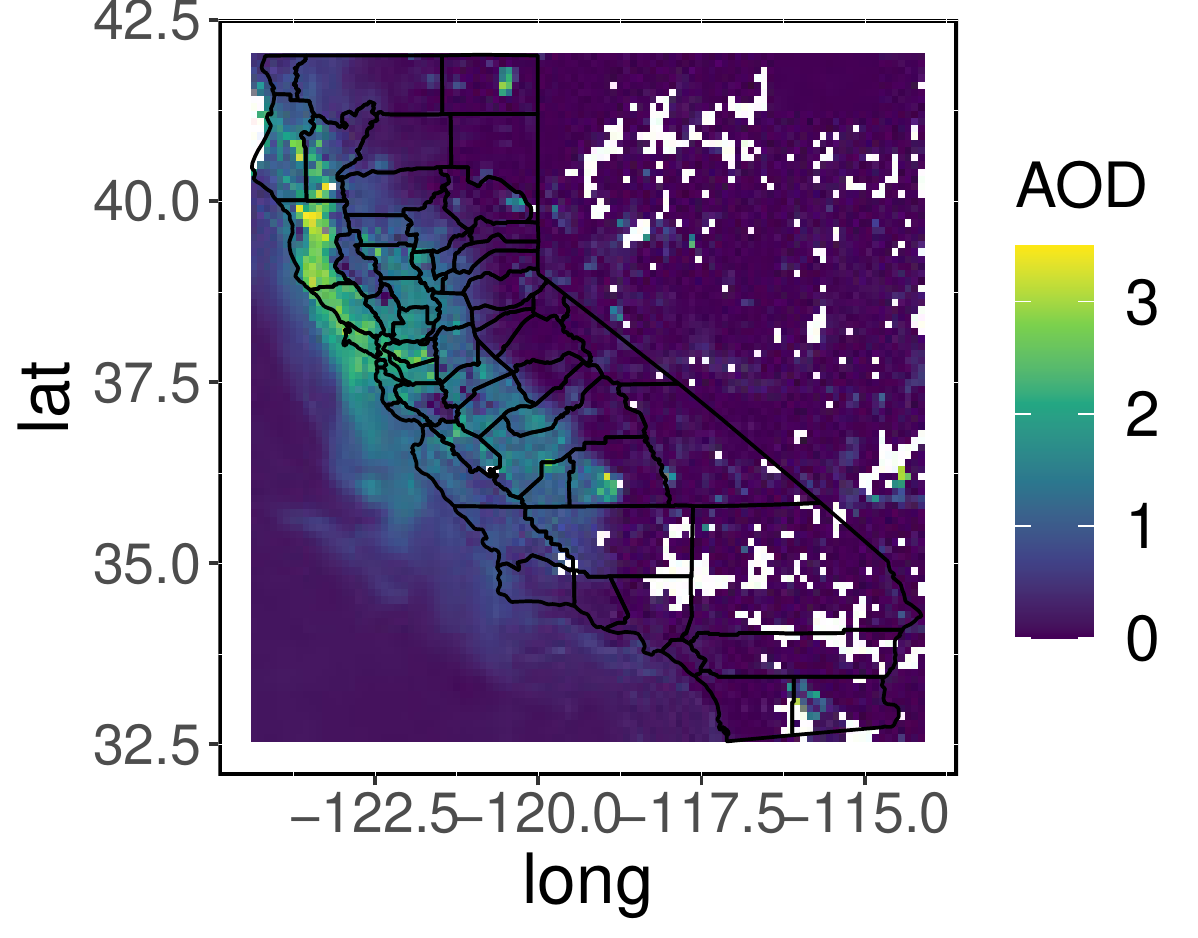}
	\end{subfigure}
	\quad
	\begin{subfigure}[t]{0.31\linewidth}
		\centering
		\subcaption{2020-10-01-18:13}
		\includegraphics[scale=0.42]{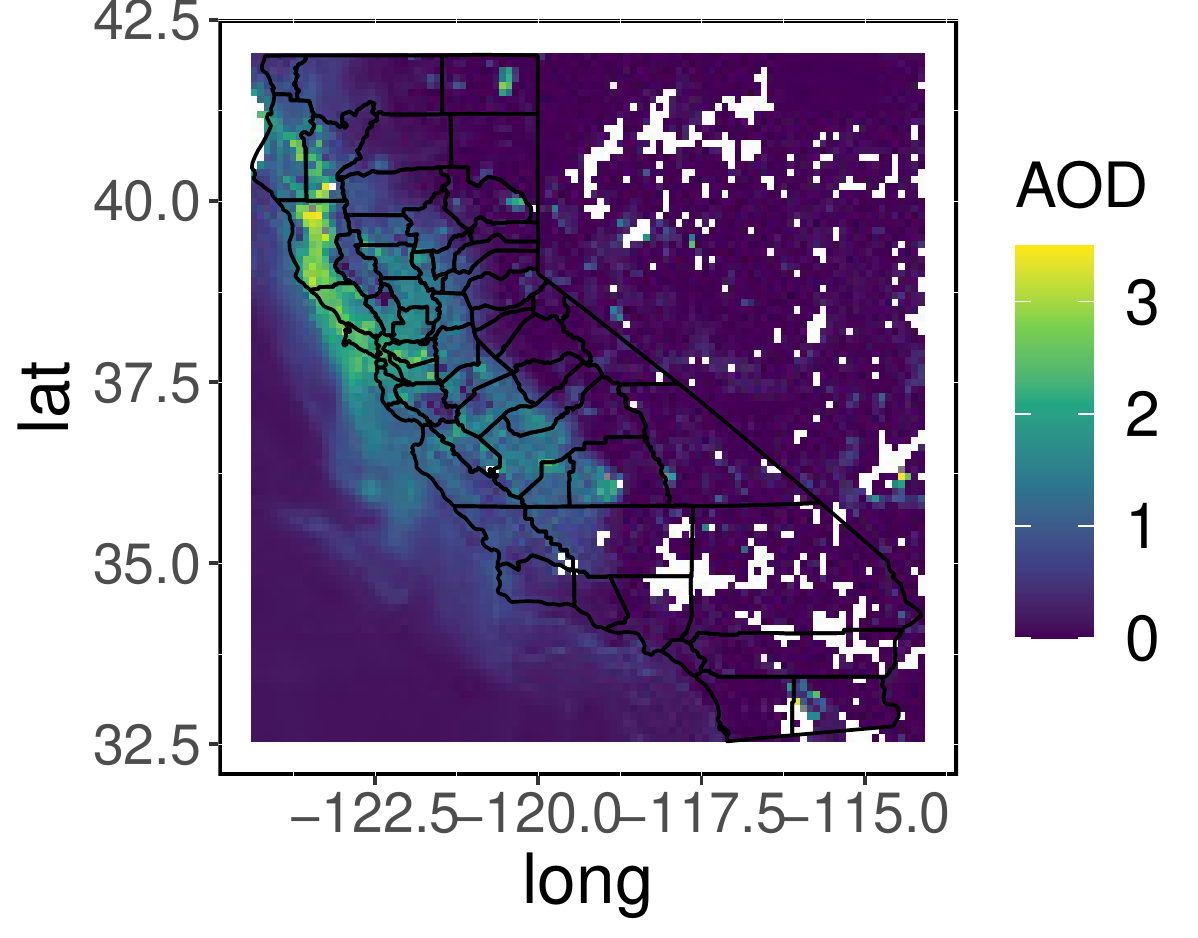}
	\end{subfigure}
	\quad
	\begin{subfigure}[t]{0.31\linewidth}
		\centering
		\subcaption{2020-10-01-18:18}
		\includegraphics[scale=0.42]{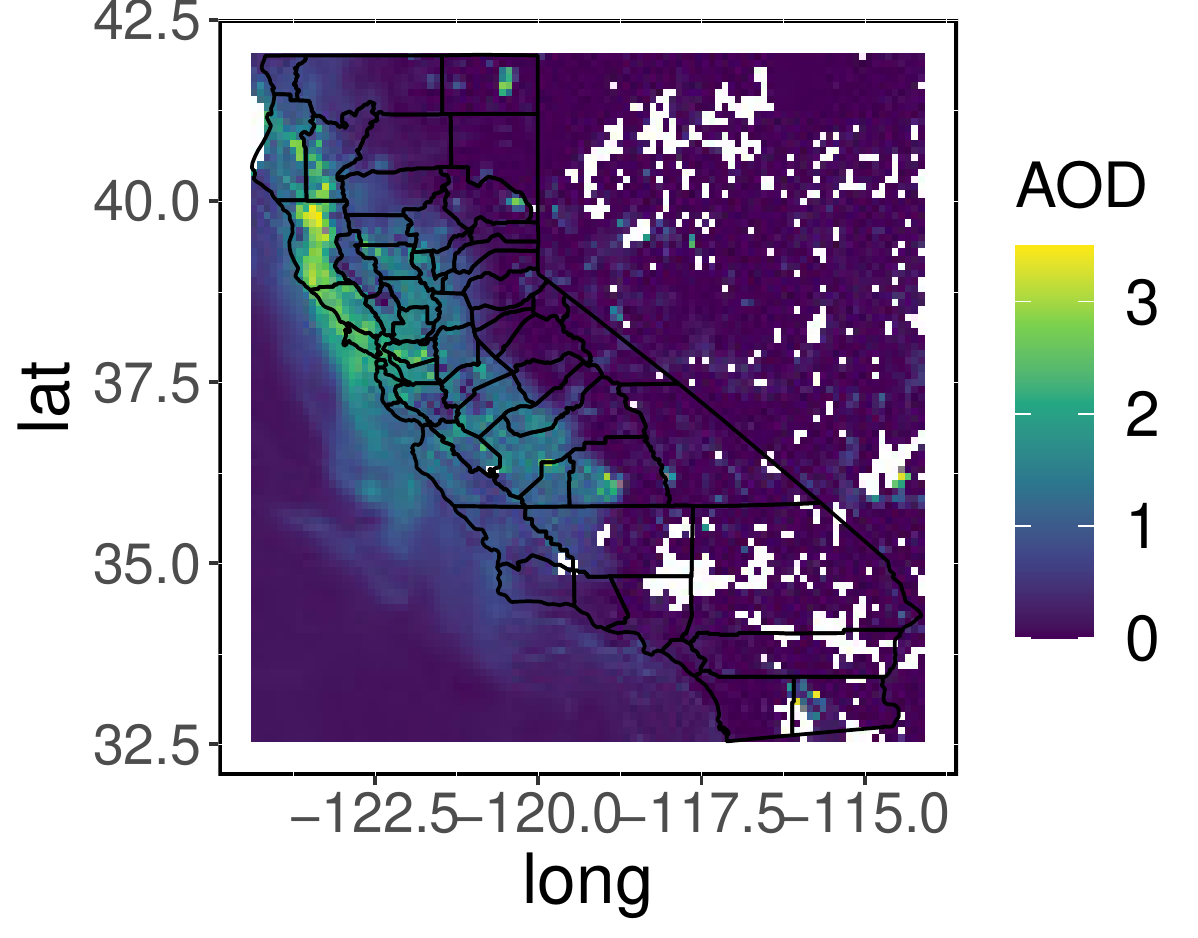}
	\end{subfigure}
	\caption{(a) Pixel centers for obtaining regular grid points; (b)-(e) Processed AOD images on October 1st 2020, from GOES-17 AOD data product.}\label{fig:Data_process_illustration}
\end{figure}

The original AOD images are available every 5 minutes and have a 56$\mu$rad$\times$56$\mu$rad
spatial resolution in scanning and sweep angles. These two axes, i.e., scanning and
sweep angles, in the raw dataset cause unevenly-spaced observational data when the AOD
data are projected to regular spatial grids with longitude and latitude. Hence, we further process the data by projecting the original data to a 0.04\textdegree$\times$0.04\textdegree \ ($\sim$20km$^2$) uniform grid as shown in Figure \ref{fig:Data_process_illustration}(a). Figure \ref{fig:Data_process_illustration} (b)-(e) show the processed AOD images on October 1, 2020, from the GOES-17 AOD dataset.

\subsection{Modeling with Data Streams from GOES-16 and GOES-17}\label{sec:result2}
\begin{figure}[h!]
    \centering
    \includegraphics[width=0.9\textwidth]{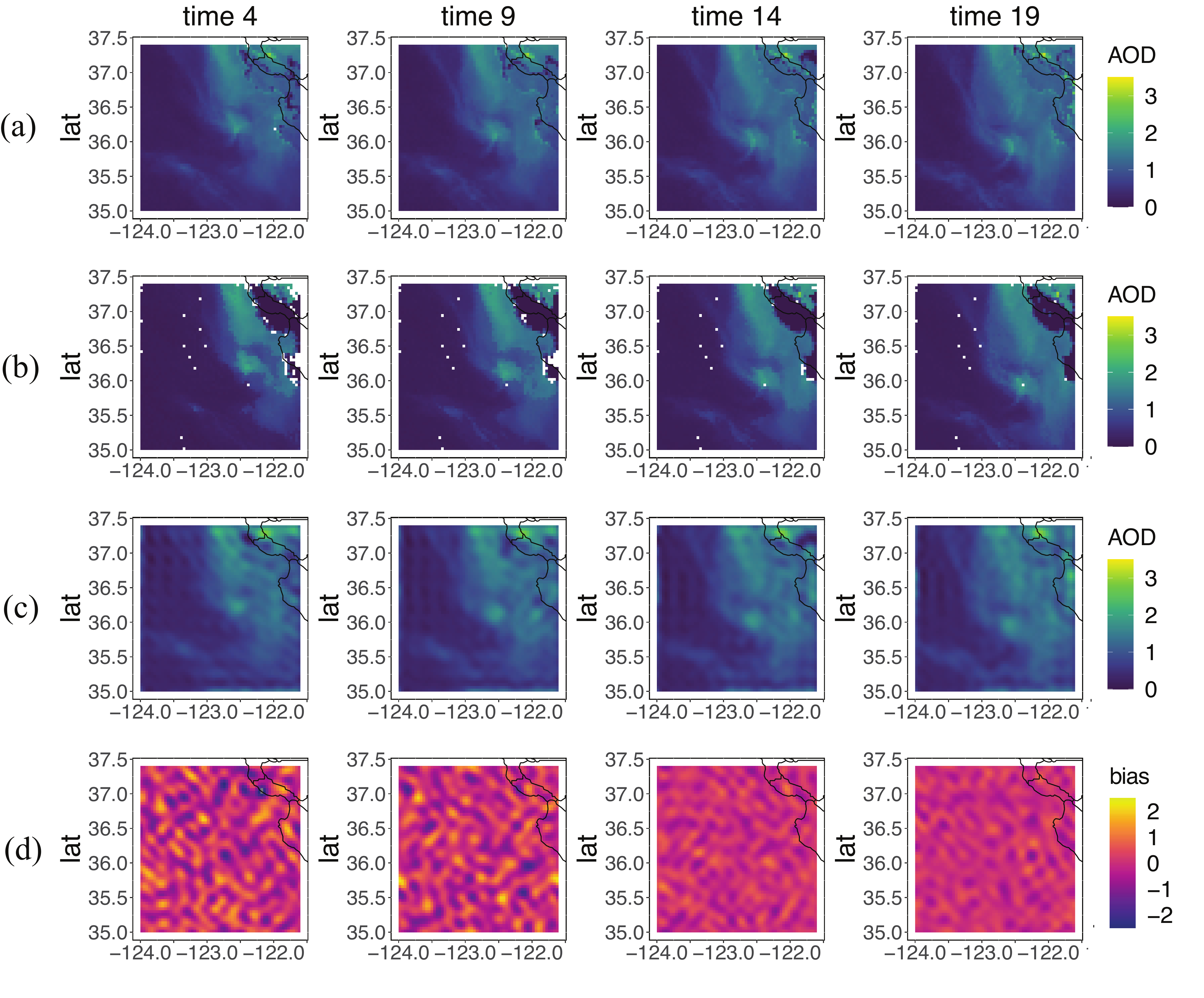}
    \caption{Row (a)-(d): GOES-17 observed AOD, GOES-16 observed AOD, filtered AOD, and the bias correction process at selected time steps $t=4,9,14$ and $19$.}\label{fig:two_source}
\end{figure}

After processing the raw data, we demonstrate the capability of the proposed model in modeling data streams from both GOES-17 and GOES-16. Figure \ref{fig:two_source} shows the AOD data stream from GOES-17 (row (a)) and GOES-16 (row (b)) at selected times at 4, 9, 14 and 19. 

As shown in the first two rows of Figure \ref{fig:two_source}, although the two image streams are obtained from the same underlying aerosols propagation process, they present very heterogeneous characteristics. For example, due to different observing angles and instruments, GOES-16 AOD images have more missing data points than GOES-17 (i.e., the white areas). More importantly, the two satellites even provide very different AOD levels in the Northeast region of the spatial domain (possibly due to measurement errors). There is a clear need to integrate, or fuse, the two heterogeneous datasets when estimating the underlying aerosols propagation. 

A number of 40 consecutive AOD images are used to establish the statistical model (20 images are respectively from GOES-16 and GOES-17). These  images are taken from 2020-10-01-18:03 to 2020-10-01-19:38 with a temporal resolution of 5 minutes. At time $t$, the data $\bm{y}_t=\text{vec}
(\bm{y}^{(1)}_t,\bm{y}^{(2)}_t)$ are arranged in a $(7200-n_t^{(1)}-n_t^{(2)})\times$1 column vector, where $n_t^{(1)}$ and $n_t^{(2)}$ are respectively the numbers of missing points from GOES-16 and GOES-17 at that time. The mapping matrices $\bm{K}^{(1)}_t$ and $\bm{K}^{(2)}_t$ are also introduced to address the missing data issue, where $\bm{K}^{(1)}_t$ is a $(3600-n_t^{(1)})\times3600$ matrix, and $\bm{K}^{(2)}_t$ is a $(3600-n_t^{(2)})\times3600$ matrix. The dynamic model for the two data streams can be written as 
\begin{equation}\label{eq:twosourcesSSM}
\begin{split}
    \begin{pmatrix} 
        \bm{y}^{(1)}_t\\\bm{y}^{(2)}_t
    \end{pmatrix} &= 
    \begin{pmatrix}
        \bm{\bm{K}^{(1)}_t F, 0}\\\bm{\bm{K}^{(2)}_t F, 0}
    \end{pmatrix} 
    \bm{\theta}_t + \bm{v}_t, \hspace{0.2in} \bm{v}_t\sim \mathcal{N}\left(\textbf{0}, \begin{bmatrix}
        \sigma_1^2\bm{I} & \\
        & \sigma_2^2\bm{I} 
    \end{bmatrix}\right)\\
    \bm{\theta}_t  &= \bm{G}_t\bm{\theta}_{t-1} + \bm{w}_{t-1}, \hspace{0.62in} \bm{w}_t\sim \mathcal{N}(\textbf{0},\bm{W}),
\end{split}
\end{equation}
where $\sigma^2_1 \sim \mathcal{IG}(a_1,b_1)$, $\sigma^2_2 \sim \mathcal{IG}(a_2,b_2)$, $\bm{W} \sim \mathcal{W}^{-1}(\bm{\Phi}, \nu)$, $a_1,b_1,a_2,b_2,\bm{\Phi}$, and $\nu$ are the hyperparameters.

\vspace{-20pt}

\begin{figure}[h!]
	\centering
	\includegraphics[width=1\textwidth]{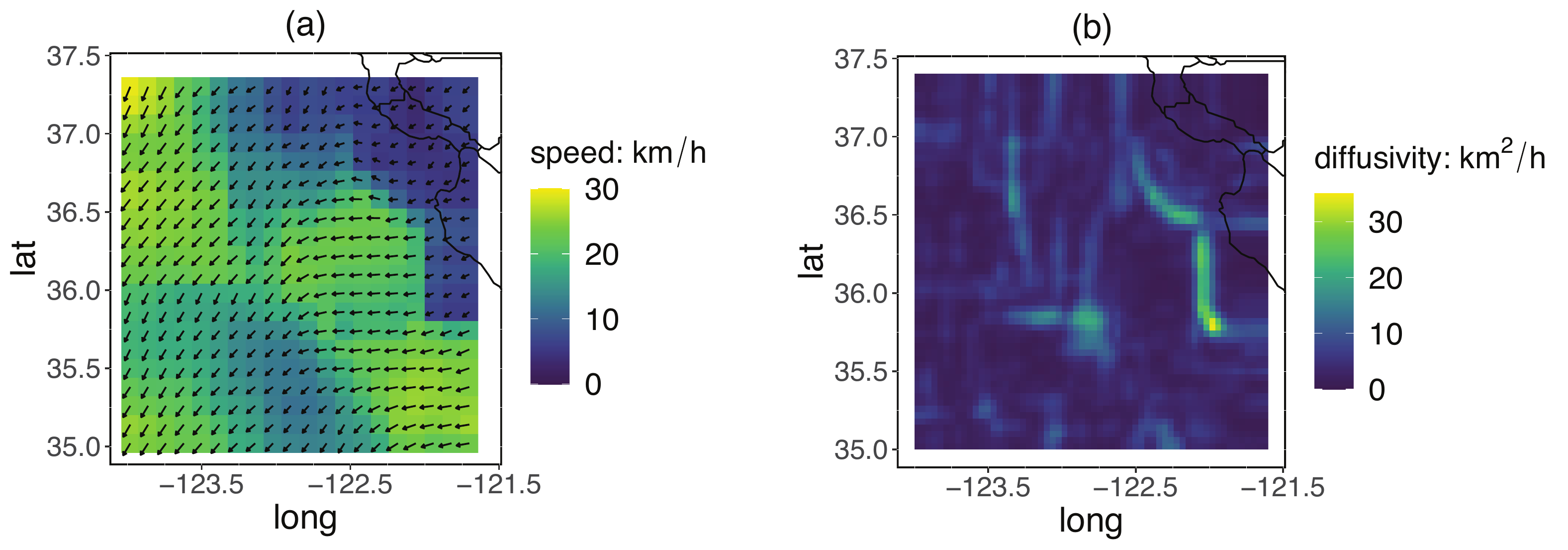}
	\caption{(a) the computed velocity field; (b) the computed diffusivity field.}\label{fig:velocity_diffusivity}
\end{figure}

Note that, the computation of $\bm{G}_t$ in (\ref{eq:twosourcesSSM}) requires the velocity field $\bm{v}(\bm{s}, t)$ and diffusivity field $\bm{D}(\bm{s}, t)$. Here, the Optical Flow (OF) method is used to obtain $\bm{v}(\bm{s}, t)$ and $\bm{D}(\bm{s},t)$. OF is a computer vision method that estimates motion vectors (velocity field in our case) from a sequence of images based on intensity gradients and differences \citep{Horn}. Figure \ref{fig:velocity_diffusivity}(a) shows the velocity field computed from the 20 AOD images using the OF method. As seen from this figure, the wind speed ranges from 3.5km/h to 29.5km/h, and the prevailing wind direction is from Northeast to Southwest. After obtaining the wind field, the diffusivity field is calculated using $\bm{D}(\bm{s})=0.28(\delta_x\delta_y)\sqrt{\left(\partial v_x\big/\partial s_x-\partial v_y\big/\partial s_y\right)^2+\left(\partial v_x\big/\partial s_y+\partial v_y\big/\partial s_x\right)^2}$, where $\delta_x$ and $\delta_y$ are the computational resolution of the wind field; see \citet{Liu2018b} for more details. Figure \ref{fig:velocity_diffusivity}(b) shows the derived diffusivity field, which ranges from 0.095km$^2$/h to 34.38km$^2$/h. 

\vspace{-12pt}

\vspace{-12pt}
Next, the Gibbs sampling with FFBS (Algorithm 1 in Appendix B) is applied to (\ref{eq:twosourcesSSM}). Figure \ref{fig:two_source} (rows (c) and (d)) show the (filtered) estimated AOD values, and the bias correction process at selected times 4, 9, 14 and 19. Here, the wavenumbers in (\ref{eq:twosourcesSSM}) are truncated at (20,20), and the augmented state is given by $\bm{\theta}_t=\text{vec}(\bm{\alpha}_t, \tilde{\bm{\gamma}}_t)\in\mathbb{R}^{800\times 1}$. We see that, 

$\diamond$ The AOD value in the Northeast region is close to 3 based on GOES-17 (the first row of Figure \ref{fig:two_source}), while the AOD value in the same region is about 0 based on GOES-16 (the second row). It is important to note that the proposed model successfully integrates (or fuses) two heterogeneous data streams in estimating the aerosols propagation, and the estimated AOD in the Northeast region is much higher than zero. The fusion of these two data images should not be seen as simply taking the average of the two data streams. Here, the estimated AOD process needs to satisfy the physical advection-diffusion process.   

$\diamond$ All missing AOD values are estimated when the estimated AOD images are reconstructed by the Inverse Fourier Transform, $\bm{F}\bm{\hat{\theta}}_t=\bm{F}\text{vec}(\hat{\bm{\alpha}}_t, \hat{\tilde{\bm{\gamma}}}_t)$.  Unlike pure data-driven missing value imputation, the missing values are computed such that the filtered AOD process satisfies the physical advection-diffusion equation.

$\diamond$ The propagation of the filtered AOD process matches the observed AOD process, which is consistent with the prevailing direction of the estimated velocity field (Northeast to Southwest). The amount of the bias correction gradually decreases, showing how the bias correction process dynamically calibrates the estimated true process.

\vspace{-8pt}
\begin{figure}[h!]
    \centering
    \includegraphics[width=1\textwidth]{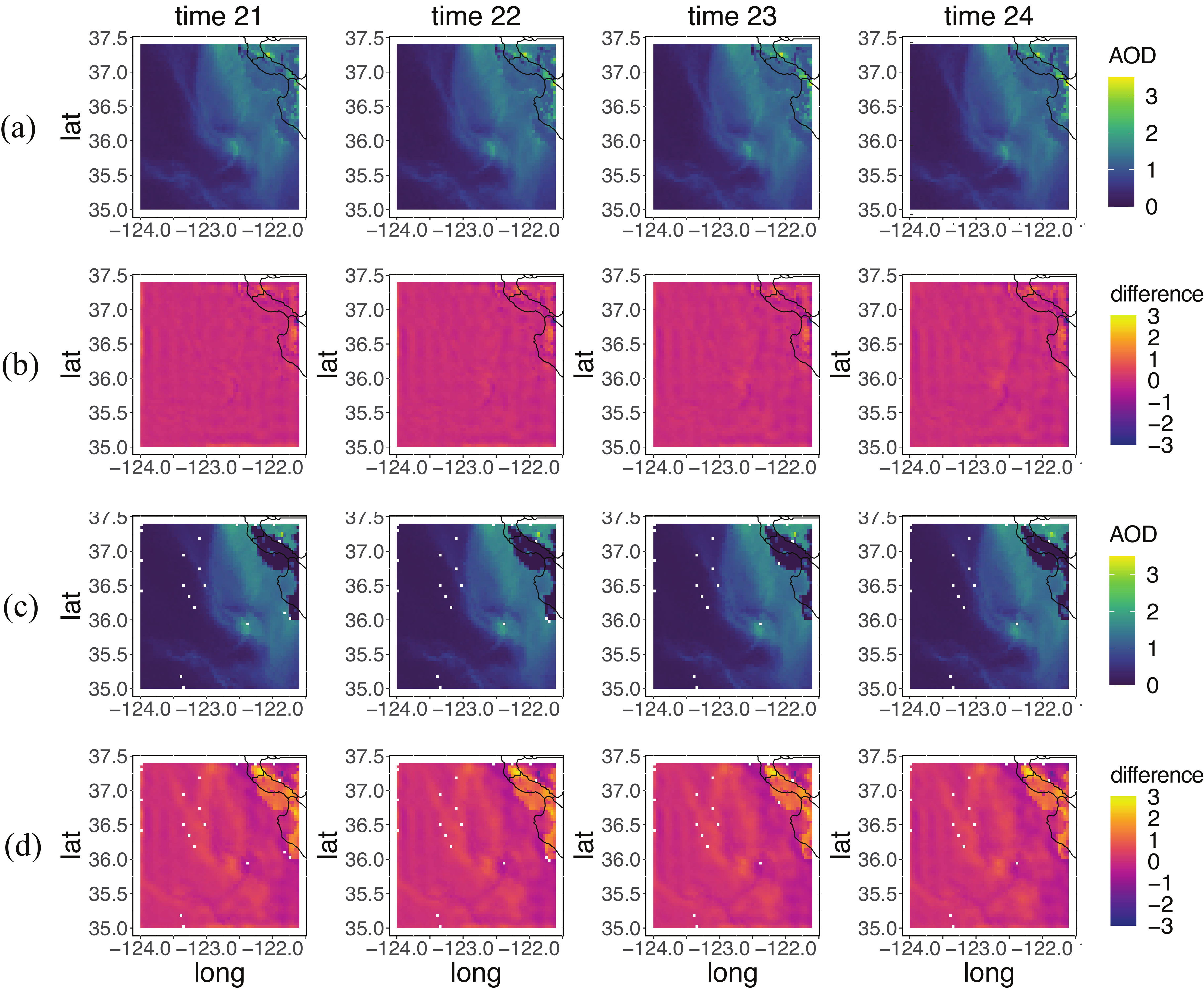}
    \caption{Row (a)-(d): observed AOD from GOES-17, difference between the estimated AOD and the observed AOD from GOES-17, observed AOD from GOES-16, and difference between the estimated AOD and the observed AOD from GOES-16 at forward time steps $t=21, 22, 23$ and 24.}\label{fig:two_source_pred}
\end{figure}

Short-term prediction of AOD is straightforward by computing the predicted state $\bm{\hat{\theta}}_{t+k}=\text{vec}(\hat{\bm{\alpha}}_{t+k}, \hat{\tilde{\bm{\gamma}}}_{t+k})$ at the forward $k$ time steps using $\bm{\hat{\theta}}_{t+k}=\bm{G}_{t+k}\bm{\hat{\theta}}_t$. Then, the predicted AOD is given by $\xi(\bm{s}, t)=\bm{F}\bm{\hat{\alpha}}_{t+k}+\bm{F}\bm{\hat{\tilde{\bm{\gamma}}}}_{t+k}$, where $\bm{F}\bm{\hat{\alpha}}_{t+k}$ and $\bm{F}\bm{\hat{\gamma}}_{t+k}$ are respectively the predicted and bias correction AOD. Figure \ref{fig:two_source_pred} shows the observed AOD values from GOES-17 (row (a)) and GOES-16 (row (c)), and the prediction error calculated based on the observations from GOES-17 (row (b)) and GOES-16 (row (c)). Note that, because both satellite data streams are possibly biased observations of the underlying AOD process, there exists no ground truth for evaluating the accuracy of the estimated AOD process. 
However, we do see from Figure \ref{fig:two_source_pred} that the prediction errors calculated using the observations from GOES-17 are less than those calculated from the GOES-16 data, implying that the predicted AOD images are more similar to GOES-17 AOD images. In addition, when the predicted horizon becomes larger, the prediction performance deteriorates as expected. The main reason is that the future velocity field gradually deviates from the velocity field estimated from historical data.

The Gibbs sampling with FFBS algorithm for fitting the model (\ref{eq:dynamicalmodel}) has cubic complexity in the number of spatial points, i.e., $\mathcal{O}(TN^3)$. Although the computational cost might be reduced by converting the data into the spectral domain by Fourier transform \citep{Sigrist}, this approach may not work well with missing data. One practical approach is to perform the data downsampling, i.e., fitting the model using a subset of selected observations (as if other data points were missing, and note that, our model can handle missing data). Hence, we investigate the trade-off between downsampling and model accuracy. 

\begin{figure}[ht]
   \begin{minipage}{0.4\textwidth}
     \centering
     \includegraphics[scale=0.5]{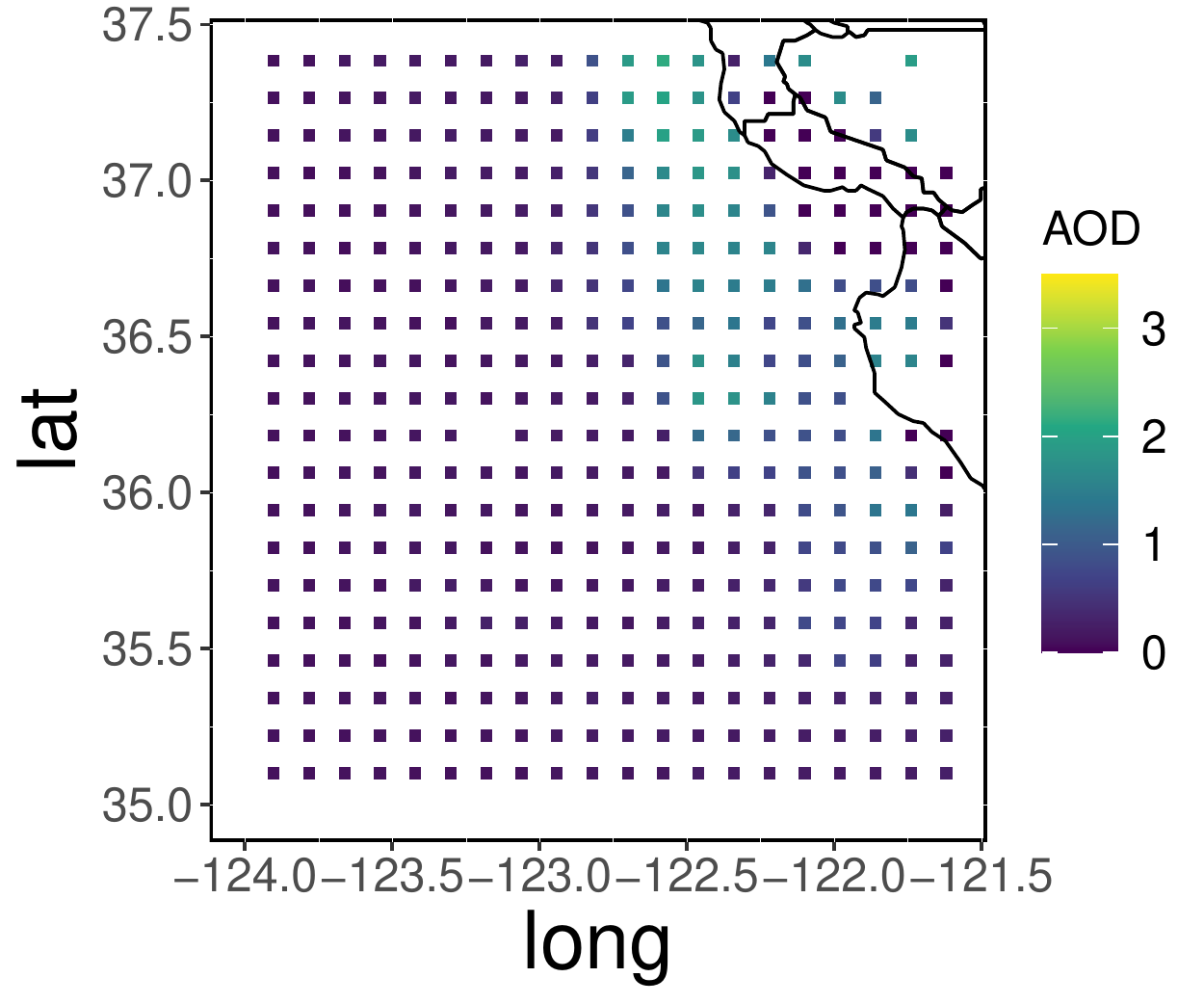}
     \caption{Uniform dowmsampling}\label{fig:uds}
   \end{minipage}\hfill
   \begin{minipage}{0.58\textwidth}
     \centering
     \includegraphics[scale=0.55]{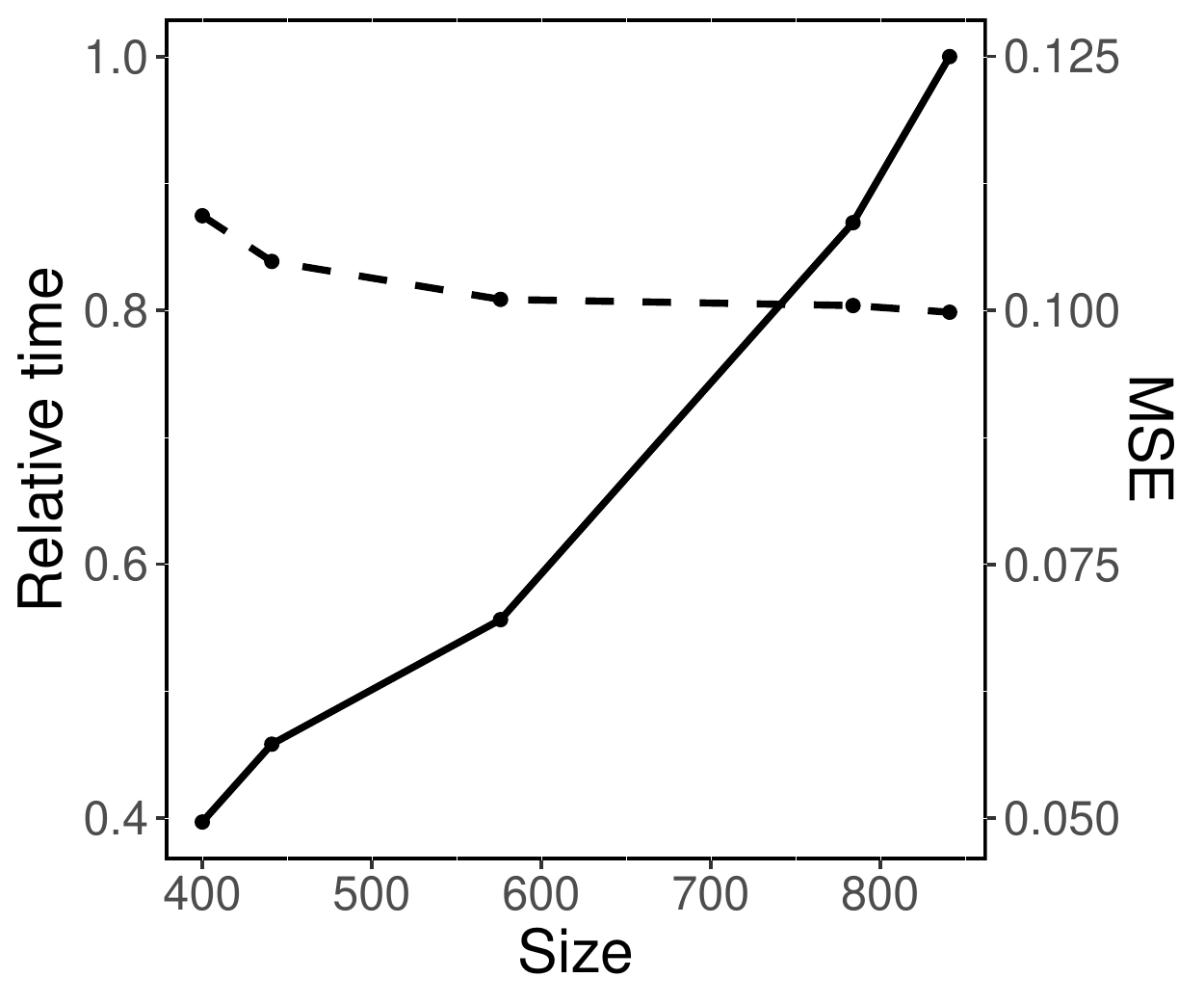}
    \caption{The trade-off between computational time and MSE as the size of the downsampling grid increases}\label{fig:computational_time}
   \end{minipage}
\end{figure}

To elaborate, we adopt the uniform downsampling as shown in Figure \ref{fig:uds}. As illustrated by this figure, only the data at the downsampling grid are used to fit the model, and the dynamic model becomes 
\begin{equation}
\begin{split}
    \bm{y}_t^{(D)} &= \bm{F}_t^{(D)}\bm{\theta}_t + \bm{v}_t^{(D)}, \hspace{1cm} \bm{v}_t\sim\mathcal{N}(\textbf{0},\bm{V}_t^{(D)})\\
    \bm{\theta}_t  &= \bm{G}_t\bm{\theta}_{t-1} + \bm{w}_{t-1}, \hspace{0.85cm} \bm{w}_t\sim \mathcal{N}(\textbf{0}, \bm{W}_t)
\end{split}
\end{equation}
where 

\begin{equation}
\bm{y}_t^{(D)} = 
  \begin{pmatrix}
    \bm{M}_t\bm{y}^{(1)}_t\\
     \bm{M}_t\bm{y}^{(2)}_t\\
    \vdots \\
    \bm{M}_t\bm{y}^{(|J_t|)}_t
   \end{pmatrix}, \quad\quad
   \bm{F}_t^{(D)} = 
    \begin{pmatrix}
    	\bm{M}_t\bm{K}^{(1)}_t \bm{F}, \bm{0}\\
    	\bm{M}_t\bm{K}^{(2)}_t \bm{F}, \bm{0}\\
    	\vdots \\
    	\bm{M}_t\bm{K}^{(|J_t|)}_t \bm{F}, \bm{0}
    \end{pmatrix}
\end{equation}
and $\bm{M}_t$ is a Boolean matrix that determines if a spatial point is sampled at time $t$.
    
In other words, the unselected points by the downsampling can be simply treated as missing data and handled by the proposed model. 
To investigate the trade-off between downsampling and model accuracy, Figure \ref{fig:computational_time} shows the relative computation times and Mean-Squared-Error (MSE) of the estimated AOD process at time 20 for various downsampling grid sizes ranging from $20\times20$ to $30\times30$. In this figure, the solid line shows the relative computation time, which grows as the downsampling grid size increases due to the cubic complexity $\mathcal{O}(TN^3)$. The dashed line shows that the estimation MSE becomes lower as the downsampling grid size becomes larger. The results clearly show that the gain in modeling accuracy decreases while the computational time significantly increases (when the downsampling grid size increases), justifying the practical need to make a trade-off between downsampling and model accuracy. 

\subsection{Comparison with the pure data-driven approach}

We further compare the proposed physics-informed model with a pure data-driven modeling approach. Note that, when the velocity field $\bm{v}(\bm{s}, t)$ and diffusivity $\bm{D}(\bm{s}, t)$ do not change over time (i.e., $\bm{G}_t$ does not vary in time in (\ref{eq:dynamicalmodel})), a pure data-driven approach can be used to estimate $\bm{G}$ as well as the system state $\bm{\theta}_t$ using the Gibbs sampling with FFBS. At each iteration of the Gibbs sampling, after obtaining $\bm{\theta}^{(i)}_{1:T}$ using FFBS, the sample $\bm{G}^{(i)}$ of $\bm{G}$ can be computed by $(\bm{\Theta}^{(i)}_1-\bm{w}^{(i-1)})(\bm{\Theta}^{(i)}_2)^{-1}$, where $\bm{\Theta}^{(i)}_1=(\bm{\theta}^{(i)}_2, \bm{\theta}^{(i)}_3, \cdots, \bm{\theta}^{(i)}_T)\in\mathbb{R}^{K\times(T-1)}$, $\bm{\Theta}^{(i)}_2=(\bm{\theta}^{(i)}_1, \bm{\theta}^{(i)}_2, \cdots, \bm{\theta}^{(i)}_{T-1})\in\mathbb{R}^{K\times(T-1)}$, $\bm{w}^{(i-1)}=(\bm{w}^{(i-1)}_1,\bm{w}^{(i-1)}_2,\cdots,\bm{w}^{(i-1)}_{T-1})$, and $\bm{w}^{(i-1)}_j \sim \mathcal{N}(\bm{0}, \bm{W}^{(i-1)})$ for $j=1,2,\cdots,T-1$. The Gibbs sampling with FFBS for the pure data-driven approach is summarized in Appendix B. 

\begin{figure}[h!]
    \centering
    \includegraphics[width=1\textwidth]{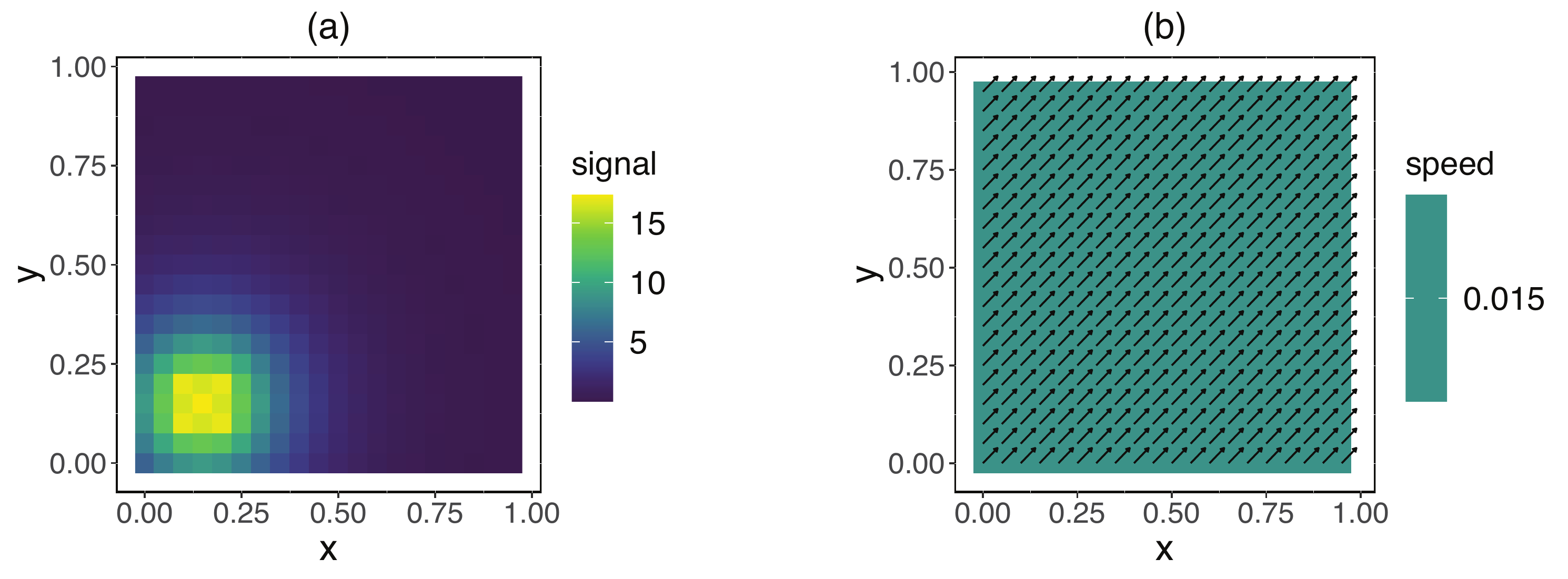}
    \caption{(a) Initial emission signals; (b) Velocity field used for simulation}\label{fig:signal}
\end{figure}
The comparison is performed with a simulated dataset based on the PDE (\ref{eq:PDE}).
As shown in Figure \ref{fig:signal}(a), an initial emission signal is generated at a 20$\times$20 spatial grid following the function:
\begin{equation}
    \begin{split}
        f(\bm{s}) =& \frac{3}{2\pi\cdot0.25^2}\exp\left(-\frac{\|\bm{s}_{\text{c1}}-\bm{s}\|^2}{2\cdot0.25^2}\right) + \frac{3}{2\pi\cdot0.25^2}\exp\left(-\frac{\|\bm{s}_{\text{c2}}-\bm{s}\|^2}{2\cdot0.25^2}\right) \\
                 &+ \frac{3}{2\pi\cdot0.25^2}\exp\left(-\frac{\|\bm{s}_{\text{c3}}-\bm{s}\|^2}{2\cdot0.25^2}\right) + \frac{3}{2\pi\cdot0.25^2}\exp\left(-\frac{\|\bm{s}_{\text{c4}}-\bm{s}\|^2}{2\cdot0.25^2}\right),
    \end{split}
\end{equation}
where $\bm{s}_{\text{c1}}=(0.1,0.1)$, $\bm{s}_{\text{c2}}=(0.1,0.2)$, $\bm{s}_{\text{c3}}=(0.2,0.1)$, and $\bm{s}_{\text{c4}}=(0.2,0.2)$ are four signal centers shown in Figure \ref{fig:signal}(a). Then, we let the initial signal propagate along the velocity field shown in Figure \ref{fig:signal}(b). Here, the velocity field has a 45\textdegree \ horizontal angle and the speed of the velocity field is 0.015 per time step over the spatial domain. A white noise process with a standard deviation of 0.1 is also included in the simulated dataset. 

A number of 30 simulated images are used to train and test the pure data-driven model. The first 20 images are used for modeling training, and the last 10 images are used for comparing the performance of the pure data-driven model and the physics-informed model. The pure data-driven model is solved using the Algorithm 2 in Appendix B. Figure \ref{fig:Data-drive-model} shows the filtered images (at times 8 and 14) and the predicted images (at times 21 and 23). Note that, the pure data-driven approach does not have a physical interpretation and the matrix $\bm{G}$ is completely estimated from the data. In addition, the wavenumber in this simulation example is truncated at $(6,6)$, and the corresponding $\bm{\theta}_t$ is a $36\times1$ column vector for the pure data-driven model. 
\begin{figure}[ht]
    \centering
    \includegraphics[width=1\textwidth]{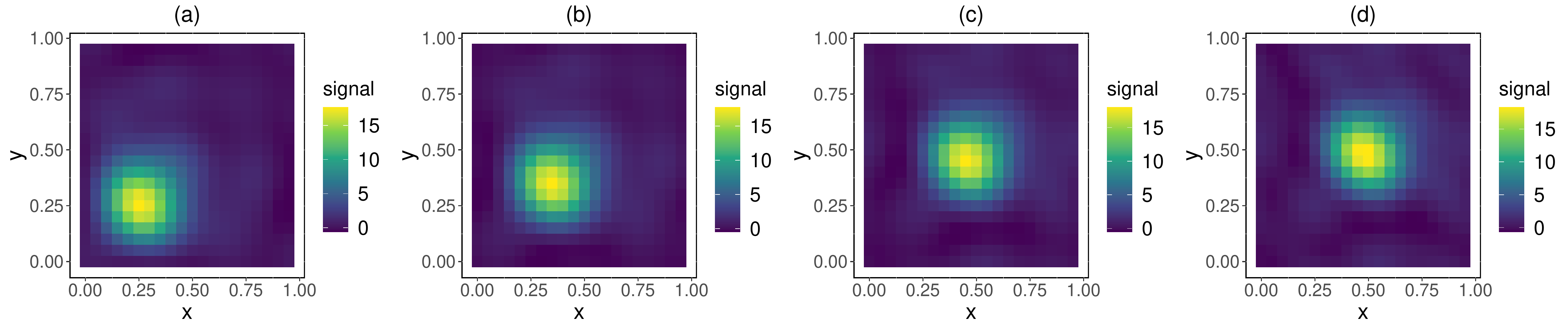}
    \caption{(a)-(b) Filtered signals at times 8 and 14; (c)-(d) Predicted signals at times 21 and 23.}\label{fig:Data-drive-model}
\end{figure}


Unlike the pure data-driven approach, the physics-informed model requires the velocity and diffusivity to be estimated before the transition matrix $\bm{G}$ can be computed. In this numerical study, we intentionally introduce errors in the specified velocity field, and investigate how the inevitable errors in practice affect the performance of the physics-informed model. In particular, we let the length of the velocity vector vary from 0.012, 0.015 (the one used in simulating the data) and 0.018, and let the direction of the velocity vector vary from 15\textdegree \ to 70\textdegree \ with 5\textdegree \ increment. The prediction MSE of the pure data-driven approach and the proposed physics-informed approach are reported in Table \ref{tab:1}.

\clearpage
\begin{table}[ht]
\caption{\label{tab:1}The prediction Mean Squared Errors (MSE) of the pure data-driven approach and the proposed physics-informed approach}
\scalebox{0.9}{
\begin{adjustbox}{angle=0} 
\begin{tabular}{c|cccccccccc} 
          \Xhline{1pt}
          \diagbox[width=7em]{Model\\Parameters}{Predicted\\MSE}& time 21 & time 22& time 23& time 24& time 25& time 26& time 27& time 28& time 29& time 30\\
          \Xhline{1pt}
          Data-driven& 0.0316& 0.0481& 0.1048& 0.2591& 0.5397& 1.0452& 1.8183& 2.9503& 4.2889& 6.0012\\
          (0.015, 45\textdegree)& 0.0284& 0.0283& 0.0250& 0.0277& 0.0293& 0.0276& 0.0297& 0.0300& 0.0306& 0.0310\\
          (0.015, 50\textdegree)& 0.0284& 0.0315& 0.0346& 0.0472& 0.0605& 0.0813& 0.1057& 0.1333& 0.1670& 0.1912\\
          (0.015, 40\textdegree)& 0.0285& 0.0293& 0.0316& 0.0443& 0.0622& 0.0735& 0.0969& 0.1236& 0.1504& 0.1936\\
          (0.015, 55\textdegree)& 0.0286& 0.0391& 0.0606& 0.1031& 0.1558& 0.2342& 0.3242& 0.4314& 0.5560& 0.6688\\
          (0.015, 35\textdegree)& 0.0288& 0.0345& 0.0543& 0.0969& 0.1588& 0.2182& 0.3059& 0.4114& 0.5225& 0.6724\\
          (0.015, 60\textdegree)& 0.0290& 0.0505& 0.1018& 0.1932& 0.3120& 0.4812& 0.6772& 0.9125& 1.1809& 1.4414\\
          (0.015, 30\textdegree)& 0.0293& 0.0436& 0.0923& 0.1836& 0.3163& 0.4573& 0.6499& 0.8832& 1.1320& 1.4462\\
          (0.015, 65\textdegree)& 0.0294& 0.0661& 0.1587& 0.3175& 0.5280& 0.8194& 1.1587& 1.5659& 2.0243& 2.4825\\
          (0.015, 25\textdegree)& 0.0298& 0.0570& 0.1464& 0.3055& 0.5345& 0.7887& 1.1237& 1.5293& 1.9622& 2.4886\\
          (0.015, 70\textdegree)& 0.0300& 0.0855& 0.2302& 0.4738& 0.7996& 1.2410& 1.7563& 2.3718& 3.0567& 3.7505\\
          (0.015, 20\textdegree)& 0.0304& 0.0739& 0.2143& 0.4580& 0.8066& 1.2023& 1.7122& 2.3274& 2.9819& 3.7560\\
          (0.015, 75\textdegree)& 0.0308& 0.1086& 0.3156& 0.6600& 1.1225& 1.7382& 2.4564& 3.3089& 4.2463& 5.2001\\
          (0.015, 15\textdegree)& 0.0312& 0.0944& 0.2958& 0.6400& 1.1293& 1.6906& 2.4022& 3.2557& 4.1578& 5.2003\\
          (0.012, 45\textdegree)& 0.0295& 0.0408& 0.0805& 0.1487& 0.2409& 0.3456& 0.4805& 0.6572& 0.8115& 1.0226\\
          (0.012, 50\textdegree)& 0.0295& 0.0435& 0.0886& 0.1647& 0.2661& 0.3886& 0.5412& 0.7386& 0.9187& 1.1484\\
          (0.012, 40\textdegree)& 0.0295& 0.0414& 0.0856& 0.1617& 0.2667& 0.3810& 0.5323& 0.7294& 0.9033& 1.1460\\
          (0.012, 55\textdegree)& 0.0297& 0.0499& 0.1098& 0.2098& 0.3427& 0.5111& 0.7154& 0.9743& 1.2250& 1.5226\\
          (0.012, 35\textdegree)& 0.0298& 0.0458& 0.1044& 0.2046& 0.3451& 0.4975& 0.6993& 0.9581& 1.1967& 1.5199\\
          (0.012, 60\textdegree)& 0.0300& 0.0593& 0.1433& 0.2824& 0.4683& 0.7089& 0.9969& 1.3559& 1.7187& 2.1299\\
          (0.012, 30\textdegree)& 0.0302& 0.0535& 0.1356& 0.2749& 0.4720& 0.6885& 0.9725& 1.3310& 1.6758& 2.1236\\
          (0.012, 65\textdegree)& 0.0304& 0.0721& 0.1895& 0.3825& 0.6419& 0.9794& 1.3808& 1.8743& 2.3858& 2.9498\\
          (0.012, 25\textdegree)& 0.0308& 0.0647& 0.1797& 0.3733& 0.6477& 0.9539& 1.3503& 1.8440& 2.3323& 2.9428\\
          (0.012, 70\textdegree)& 0.0311& 0.0887& 0.2488& 0.5102& 0.8627& 1.3201& 1.8614& 2.5193& 3.2096& 3.9572\\
          (0.012, 20\textdegree)& 0.0314& 0.0793& 0.2363& 0.4982& 0.8692& 1.2879& 1.8232& 2.4823& 3.1448& 3.9474\\
          (0.012, 75\textdegree)& 0.0320& 0.1087& 0.3201& 0.6632& 1.1265& 1.7236& 2.4273& 3.2735& 4.1654& 5.1181\\
          (0.012, 15\textdegree)& 0.0322& 0.0962& 0.3026& 0.6451& 1.1293& 1.6805& 2.3768& 3.2260& 4.0861& 5.1009\\
          (0.018, 45\textdegree)& 0.0306& 0.0500& 0.0860& 0.1559& 0.2579& 0.3837& 0.5345& 0.6893& 0.8998& 1.0735\\
          (0.018, 50\textdegree)& 0.0306& 0.0535& 0.0968& 0.1782& 0.2940& 0.4458& 0.6223& 0.8087& 1.0568& 1.2565\\
          (0.018, 40\textdegree)& 0.0308& 0.0514& 0.0940& 0.1758& 0.2966& 0.4381& 0.6137& 0.7986& 1.0397& 1.2631\\
          (0.018, 55\textdegree)& 0.0306& 0.0614& 0.1255& 0.2415& 0.4026& 0.6209& 0.8728& 1.1510& 1.5029& 1.8023\\
          (0.018, 35\textdegree)& 0.0310& 0.0565& 0.1191& 0.2352& 0.4068& 0.6048& 0.8554& 1.1313& 1.4697& 1.8162\\
          (0.018, 60\textdegree)& 0.0307& 0.0741& 0.1728& 0.3458& 0.5838& 0.9079& 1.2826& 1.7086& 2.2246& 2.6901\\
          (0.018, 30\textdegree)& 0.0311& 0.0664& 0.1624& 0.3356& 0.5891& 0.8826& 1.2551& 1.6784& 2.1747& 2.7084\\
          (0.018, 65\textdegree)& 0.0308& 0.0909& 0.2369& 0.4883& 0.8319& 1.2973& 1.8368& 2.4593& 3.1897& 3.8752\\
          (0.018, 25\textdegree)& 0.0313& 0.0808& 0.2234& 0.4752& 0.8401& 1.2651& 1.8018& 2.4221& 3.1272& 3.8993\\
          (0.018, 70\textdegree)& 0.0309& 0.1126& 0.3188& 0.6689& 1.1453& 1.7836& 2.5243& 3.3827& 4.3655& 5.3092\\
          (0.018, 20\textdegree)& 0.0316& 0.0999& 0.3019& 0.6526& 1.1559& 1.7447& 2.4818& 3.3396& 4.2920& 5.3376\\
          (0.018, 75\textdegree)& 0.0312& 0.1377& 0.4154& 0.8824& 1.5154& 2.3538& 3.3249& 4.4492& 5.7096& 6.9342\\
          (0.018, 15\textdegree)& 0.0320& 0.1224& 0.3946& 0.8619& 1.5273& 2.3068& 3.2737& 4.3998& 5.6259& 6.9638\\
          \Xhline{1pt}
\end{tabular} 
\end{adjustbox}}
\end{table}

\begin{figure}[h!]
	\centering
    \includegraphics[width=1\textwidth]{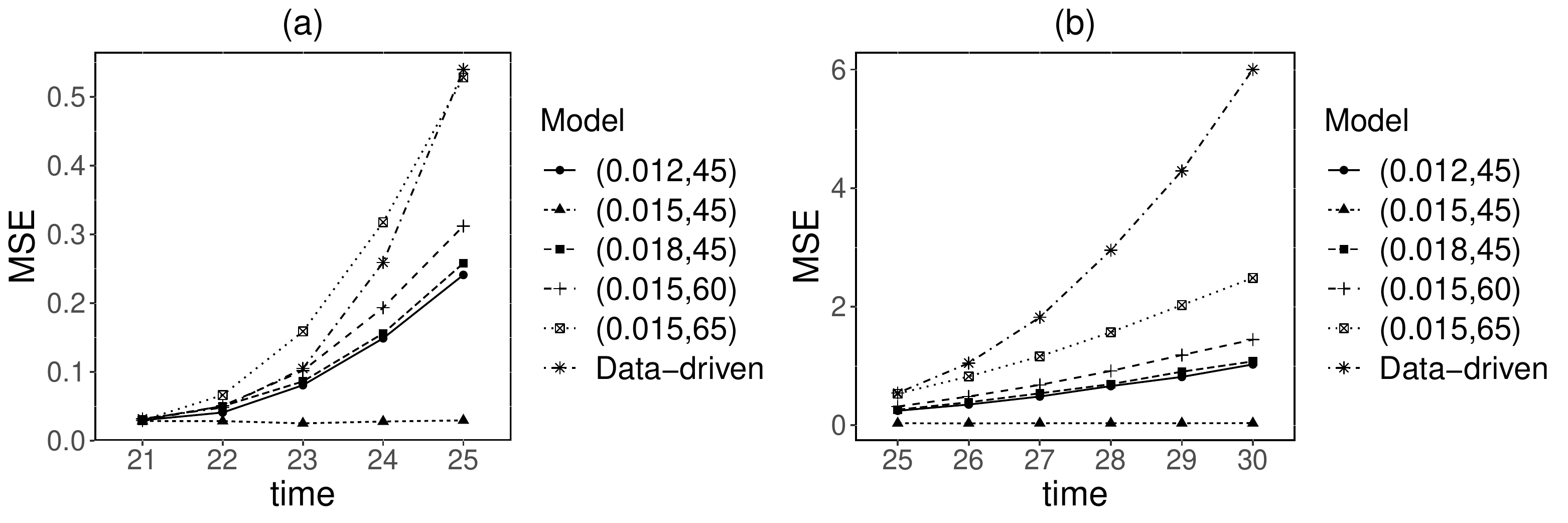}
	\caption{(a) Prediction MSEs at forward times $t=21, 22, 23, 24, 25$; (b) Prediction MSEs at forward times $t=25, 26, 27, 28, 29, 30$.}\label{fig:comparision}
\end{figure}

To better visualize the results reported in Table 1, Figure \ref{fig:comparision} shows the prediction MSEs for the pure data-driven model and 5 selected physics-informed models. For example, $(0.012, 45)$ denotes the physics-informed model based on a pre-specified velocity field with a speed of 0.012 per time unit and a horizontal angle of 45\textdegree. Some important observations are summarized as follows:

$\bullet$ As the prediction horizon grows, the physics-informed model based on the correctly pre-specified velocity field (i.e., the ideal but less realistic case in practice) consistently has the lowest prediction MSE (i.e., dashed line with $\blacktriangle$). 

$\bullet$ Even if the pre-specified velocity fields deviate away from the actual velocity field used for simulating the data, the physics-informed model can still produce lower MSEs under most of the comparison scenarios. For example, as shown in Figure \ref{fig:comparision}(a) and (b), the MSEs of the physics-informed models (lines with $+$, $\blacksquare$ and $\bullet$) have lower MSEs than the pure data-driven model. These three physics-informed models are based on the pre-specified speed which is deviated from the true value. On the other hand, for the physics-informed model based on the pre-specified velocity information $(0.015,65)$ (i.e., the pre-specified direction is 20\textdegree \ higher than the true degree), the prediction MSEs of the physics-informed model are larger than that of the data-driven approach from times 21 to 25 (Figure \ref{fig:comparision}(a)). It is worth noting that, when the prediction horizon further extends (Figure \ref{fig:comparision}(b)), the prediction MSEs of this physics-informed model become lower than that of the pure data-driven model over times 26 to 30. This observation strongly suggests the enhanced predictive capability of the physics-informed statistical model over pure data-driven approaches for extrapolation in time, as long as the estimated physical parameters are not too far away from the true values. 

\section{Conclusions}
The paper proposed a physics-informed spatio-temporal model for wildfire aerosols propagation using multi-source remote-sensing data streams. The model has successfully integrated multi-source remote-sensing data streams with the underlying physical advection-diffusion process, and is able to handle the heterogeneity among multi-source data streams (e.g., missing data, systematic biases, etc.). The bias correction term enables the model to account for the inadequacy of the first-order governing physics as well as the bias due to the truncation of the Fourier series. The Gibbs sampling with FFBS has been used to estimate the unknown parameters and system state variables. The proposed approach has been successfully applied to model the AOD process for the 2020 Glass Fire in California utilizing both the GOES-16 and GOES-17 satellite data streams. The modeling and short-term predictive capabilities of the proposed model have been demonstrated. Comprehensive numerical investigations have shown the advantages of the proposed physics-informed statistical model over pure data-driven alternatives. Finally, it is worth pointing out that, the proposed method for modeling wildfire aerosols propagation can be applied to other advection-diffusion types of environmental and engineering processes with multi-source observations (e.g. sea surface temperature, air pollution, and material corrosion and contamination, etc.). 
Computer code is available at: \url{https://github.com/gz-wei/Physics-Informed-Statistical-Modeling-for-Wildfire-Aerosols-Propagation}.

%
%
%
\section*{Appendix A}
Submitted as supplemental materials. 

\section*{Appendix B}\label{sec:Algorithm2}
Submitted as  supplemental materials. 

\bibliographystyle{nonumber}

\begin{thebibliography}{}

\bibitem[{Banerjee et al(2014)}]{Banerjee}
Banerjee, S., Carlin, B. P., and Gelfand, A. E. (2014). \emph{Hierarchical  modeling  and  analysis for spatial data}. CRC press.

\bibitem[{Cressie and Wikle(2015)}]{Cressie}
Cressie, N. and Wikle, C. K. (2015). \emph{Statistics for spatio-temporal data}. John Wiley \& Sons.

\bibitem[{Cushman-Roisin and Beckers(2011)}]{Cushman}
Cushman-Roisin, B. and Beckers, J.-M. (2011). \emph{Introduction to geophysical fluid dynamics: physical and numerical aspects}. Academic press.

\bibitem[{Ezzat et al.(2019)}]{Ezzat}
Ezzat, A. A., Jun, M., and Ding, Y. (2019). Spatio-temporal short-term wind forecast: A calibrated regime-switching method. \emph{The Annals of Applied Statistics}, 13(3):1484.

\bibitem[{Fang et al.(2019)}]{Fang}
Fang, X., Paynabar, K., and Gebraeel, N. (2019). Image-based prognostics using penalized tensor regression. \emph{Technometrics}, 61(3):369–384.

\bibitem[{Gibbs(1898)}]{Gibbs}
Gibbs, J. W. (1898). Fourier's Series, https://doi.org/10.1038/059200b0.


\bibitem[{Google Cloud(2021)}]{Google}
Google Cloud (2021). GOES-16 dataset. https://console.cloud.google.com/storage/browser/gcp-public-data-goes-16.

\bibitem[{Guinness and Stein(2013)}]{Guinness}
Guinness, J. and Stein, M. L. (2013). Interpolation of nonstationary high frequency spatial–temporal temperature data. \emph{The Annals of Applied Statistics}, 7(3):1684–1708.

\bibitem[{Horn and Schunck(1981)}]{Horn}
Horn, B. K. and Schunck, B. G. (1981). Determining optical flow. \emph{Artificial intelligence}, 17(1-3):185–203.

\bibitem[{Kang and Cressie(2011)}]{Kang}
Kang, E. L. and Cressie, N. (2011). Bayesian inference for the spatial random effects model. \emph{Journal of the American Statistical Association}, 106(495):972–983.

\bibitem[{Katzfuss and Cressie(2011)}]{Katzfuss2011}
Katzfuss, M. and Cressie, N. (2011). Spatio-temporal smoothing and EM estimation for massive remote-sensing data sets. \emph{Journal of Time Series Analysis}, 32(4):430–446.

\bibitem[{Katzfuss et al.(2020)}]{Katzfuss2020} 
Katzfuss, M., Stroud, J. R., and Wikle, C. K. (2020). Ensemble kalman methods for high-dimensional hierarchical dynamic space-time models. \emph{Journal of the American Statistical Association}, 115(530):866–885

\bibitem[{Kutz et al.(2016)}]{Kutz}
Kutz, N., Brunton, S., Brunton, B., and Proctor, J. (2016). \emph{Dynamic Mode Decomposition: Data-Driven Modeling of Complex Systems}. Society for Industrial and Applied Mathematics.

\bibitem[{Kuusela and Stein(2018)}]{Kuusela}
Kuusela, M. and Stein, M. L. (2018). Locally stationary spatio-temporal interpolation of Argo profiling float data. \emph{Proceedings of the Royal Society A}, 474(2220):20180400.

\bibitem[{Liu et al.(2018a)}]{Liu2018a}
Liu, X., Yeo, K., and Kalagnanam, J. (2018a). A statistical modeling approach for spatio-temporal degradation data. \emph{Journal of Quality Technology}, 50(2):166–182.

\bibitem[{Liu et al.(2016)}]{Liu2016}
Liu, X., Yeo, K., Hwang, Y., Singh, J., and Kalagnanam, J. (2016). A statistical modeling approach for air quality data based on physical dispersion processes and its application to ozone modeling. \emph{The Annals of Applied Statistics}, 10(2):756–785.

\bibitem[{Liu et al.(2018b)}]{Liu2018b}
Liu, X., Gopal, V., and Kalagnanam, J. (2018b). A spatio-temporal modeling framework for weather radar image data in tropical Southeast Asia. \emph{The Annals of Applied Statistics}, 12(1):378–407.

\bibitem[{Liu et al.(2021)}]{Liu2021}
Liu, X., Yeo, K., and Lu, S. (2021). Statistical modeling for spatio-temporal data from stochastic convection-diffusion processes. \emph{Journal of the American Statistical Association}, available on-line: https://doi.org/10.1080/01621459.2020.1863223.

\bibitem[{Ma and Kang.(2020a)}]{Ma2020a}
Ma, P. and Kang, E. L. (2020a). A fused Gaussian process model for very large spatial data. \emph{Journal of Computational and Graphical Statistics}, 29(3):479–489.

\bibitem[{Ma and Kang.(2020b)}]{Ma2020b}
Ma, P. and Kang, E. L. (2020b). Spatio-temporal data fusion for massive sea surface temperature data from MODIS and AMSR-E instruments. \emph{Environmetrics}, 31(2):e2594.

\bibitem[{Nguyen et al.(2012)}]{Nguyen2012}
Nguyen, H., Cressie, N., and Braverman, A. (2012). Spatial statistical data fusion for remote sensing applications. \emph{Journal of the American Statistical Association}, 107(499):1004–1018.

\bibitem[{Nguyen et al.(2014)}]{Nguyen2014}
Nguyen, H., Katzfuss, M., Cressie, N., and Braverman, A. (2014). Spatio-temporal data fusion for very large remote sensing datasets. \emph{Technometrics}, 56(2):174–185.

\bibitem[{Sigrist et al.(2015)}]{Sigrist}
Sigrist F., Kunsch, H. R., and Stahel, W. A. (2015). Stochastic partial differential equation based modelling of large space–time data sets. \emph{Journal  of  the  Royal  Statistical  Society: Series B}, pages 3–33.

\bibitem[{Stein(2012)}]{Stein}
Stein, M. L. (2012). \emph{Interpolation of spatial data:  some theory for kriging}. Springer Science \& Business Media.



\bibitem[{Stroud et al.(2001)}]{Stroud2001}
Stroud, J. R., Muller, P., and Sanso, B. (2001). Dynamic models for spatiotemporal data. \emph{Journal of the Royal Statistical Society: Series B}, 63(4):673–689.

\bibitem[{Stroud et al.(2010)}]{Stroud2010}
Stroud, J. R., Stein, M. L., Lesht, B. M., Schwab, D. J., and Beletsky, D. (2010). An ensemble kalman filter and smoother for satellite data assimilation. \emph{Journal of the American Statistical Association}, 105(491):978–990.

\bibitem[{Stroud et al.(2017)}]{Stroud2017}
Stroud, J. R., Stein, M. L., and Lysen, S. (2017). Bayesian and maximum likelihood estimation for Gaussian processes on an incomplete lattice. \emph{Journal of Computational and Graphical Statistics}, 26(1):108–120.

\bibitem[{Wikle(2019)}]{Wikle2019}
Wikle, C. K. (2019). Comparison of deep neural networks and deep hierarchical models for spatio-temporal data. \emph{Journal of Agricultural, Biological and Environmental Statistics}, 24(2):175–203

\bibitem[{Wikle et al.(1998)}]{Wikle1998}
Wikle, C. K., Berliner, L. M., and Cressie, N. (1998). Hierarchical Bayesian space-time models. \emph{Environmental and Ecological Statistics}, 5(2):117–154.

\bibitem[{Wikle et al.(2001)}]{Wikle2001}
Wikle, C. K., Milliff, R. F., Nychka, D., and Berliner, L. M. (2001).  Spatiotemporal hierarchical Bayesian modeling tropical ocean surface winds. \emph{Journal of the American Statistical Association}, 96(454):382–397.

\bibitem[{Yan et al.(2018)}]{Yan}
Yan, H., Paynabar, K., and Shi, J. (2018). Real-time monitoring of high-dimensional functional data streams via  spatio-temporal smooth sparse decomposition. \emph{Technometrics}, 60(2):181–197.

\bibitem[{Yang and Qiu.(2018)}]{Yang}
Yang, K. and Qiu, P. (2018). Spatiotemporal incidence rate data analysis by nonparametric regression. \emph{Statistics in Medicine}, 37(13):2094–2107.

\bibitem[{Yao et al.(2017)}]{Yao}
Yao, B., Zhu, R., and Yang, H. (2017). Characterizing the location and extent of myocardial infarctions with inverse ECG modeling and spatiotemporal regularization. \emph{IEEE Journal of Biomedical and Health Informatics}, 22(5):1445–1455.

\bibitem[{York(2020)}]{York}
York, S. (2020). \emph{U.S. Energy Information Administration}. https://www.eia.gov/todayinenergy/detail.php?id=45336. 


\end{thebibliography}

\end{document}